\documentclass[lettersize,journal]{IEEEtran}
\usepackage{amsmath,amsfonts}
\usepackage{algorithmic}
\usepackage{algorithm}
\usepackage{array}
\usepackage[caption=false,font=normalsize,labelfont=sf,textfont=sf]{subfig}
\usepackage{textcomp}
\usepackage{stfloats}
\usepackage{url}
\usepackage{verbatim}
\usepackage{graphicx}
\usepackage{cite}
\usepackage{amssymb}  
\usepackage{bm}      
\usepackage{multirow} 
\usepackage{amsmath}
\usepackage{caption}
\usepackage{graphicx}
\usepackage{epstopdf}
\usepackage{afterpage}
\usepackage{xcolor}

\begin{document}

\title{Energy Efficient and Balanced Task Assignment Strategy for Multi-UAV Patrol Inspection System in Mobile Edge Computing Network}

\author{Kuan Jia, Dingcheng Yang, Yapeng Wang, Tianyun Shui, Chenji Liu
\thanks{This work was supported in part by the National Natural Science Foundation of China under Grant 62261035, Grant 62461040 and Grant 62061027, in
part by Jiangxi Science Technology Foundation under Grant 20223BCJ25016,Grant 20213AAE01007, Grant 20224BBC3100 and Grant 20223BBE51035. An earlier version of this paper was presented in part at the 2024 IEEE/CIC ICCC Workshops, Hangzhou, China, Aug 2024 [DOI:10.1109/ICCCWorkshops62562.2024.10693809](Corresponding author: Dingcheng Yang)}       
\thanks{K. Jia and D. Yang are with the Information Engineering School, Nanchang University, Nanchang 330031, China (Corresponding author: Dingcheng Yang, e-mail: jiakuan@email.ncu.edu.cn; yangdingcheng@ncu.edu.cn).}
\thanks{Y. Wang is with the Applied Sciences School, Macao Polytechnic University, Macao 999078, China (e-mail: yapengwang@mpu.edu.mo)}
\thanks{T. Shui and C. Liu are with China Mobile Communications Group Jiangxi Co., Ltd. Nanchang 330029, China (e-mail: shuitianyun@jx.chinamobile.com; liuchenji@jx.chinamobile.com)}
}

\maketitle

\setlength{\abovecaptionskip}{1pt plus 1pt minus 1pt} 
\setlength{\belowcaptionskip}{1pt plus 1pt minus 1pt} 

\begin{abstract}
  This paper considers a patrol inspection scenario where multiple unmanned aerial vehicles (UAVs) are adopted to traverse multiple predetermined cruise points for data collection. The UAVs are connected to cellular networks and they would offload the collected data to the ground base stations (GBSs) for data processing within the constrained duration. This paper proposes a balanced task assignment strategy among patrol UAVs and an energy-efficient trajectory design method. Through jointly optimizing the cruise point assignment, communication scheduling, computational allocation, and UAV trajectory, a novel solution can be obtained to balance the multiple UAVs' task completion time and minimize the total energy consumption. Firstly, we propose a novel clustering method that considers geometry topology, communication rate, and offload volume; it can determine each UAV's cruise points and balance the UAVs' patrol task. Secondly, a hybrid Time-Energy traveling salesman problem is formulated to analyze the cruise point traversal sequence, and the energy-efficient UAV trajectory can be designed by adopting the successive convex approximation (SCA) technique and block coordinate descent (BCD) scheme. The numerical results demonstrate that the proposed balanced task assignment strategy can efficiently balance the multiple UAVs' tasks. Moreover, the min-max task completion time and total energy consumption performance of the proposed solution outperform that of the current conventional approach.
\end{abstract}   

\section{Introduction}

\IEEEPARstart{T}{he} Unmanned Aerial Vehicle (UAV) has emerged as a powerful tool in the low-altitude economy. Thanks to its exceptional mobility, flexibility, and scalability, UAVs can be deployed in various scenarios such as aerial photography, emergency rescue, and tracking. By equipping UAVs with communication devices, they can integrate with cellular networks, thereby enhancing their operational capabilities and ensuring safety \cite{ref1}. Therefore, UAVs hold great promise as a valuable asset in large-scale patrol inspection systems.

Recent research has optimized UAV trajectories and power management to enhance energy efficiency and communication performance in wireless networks. The work in \cite{pan2023joint} proposed an innovative approach for integrating power control with 3D trajectory optimization in UAV-enabled wireless-powered communication networks, significantly improving energy utilization and overcoming environmental obstacles. Similarly, the authors in \cite{liu2021joint} introduced a pioneering strategy for joint scheduling and trajectory optimization of charging UAVs in wireless rechargeable sensor networks, improving charging efficiency by reducing both hovering points and flight distance.

However, due to the limited payload capacity and endurance of UAVs, they cannot be equipped with powerful computational devices to handle resource-intensive tasks individually, such as lidar signal processing and high-definition video analysis. To address this challenge, the UAV patrol inspection system utilizes mobile edge computing (MEC) technology, which leverages computing resources deployed at the ground base station. By integrating cellular-connected UAVs with MEC, the patrol inspection system can efficiently manage complex tasks such as power inspection, military reconnaissance, cargo delivery, and remote sensing across a wide area.

For the purpose of accomplishing the implementation of extensive inspection tasks in a more efficient and quicker manner, multi-UAV collaboration is adopted to conduct the inspection operations for the patrol inspection system. Due to the diversity of inspection and flight tasks, how to balance the task allocation among multiple UAVs will consequently become an important issue. In this paper, we will analyze the energy-efficient and balanced task assignment strategy to optimize the total energy consumption and task completion time. It can assist multiple UAVs to accomplish complex inspection tasks within a broader scope by means of division of labor and cooperation.

\subsection{Related Work}
To maximize the benefits of UAVs in communication, researchers have devoted considerable attention to developing UAV-assisted communication systems. The authors in \cite{pan2023resource} introduce a novel optimization framework to address the complex problem of resource scheduling in UAV-aided device-to-device (D2D) networks, significantly enhancing network capacity and energy efficiency. Considering the practical challenges of imperfect channel state information (CSI) and coordinate uncertainties, the work in \cite{xu2021robust} further proposes a robust resource allocation algorithm for energy-harvesting-based D2D communication, substantially improving network robustness and efficiency under real-world conditions.

With the increasing prevalence of GBSs, integrating UAVs into cellular networks for various tasks has become a new trend. Compared with traditional UAVs operating solely as aerial base stations, cellular-connected UAVs offer significant advantages such as greater task execution range and real-time connectivity \cite{ref2}. Extensive research has explored the field of cellular-connected UAVs. For instance, the work in \cite{ref3} and \cite{ref4} focuses on optimizing communication performance and energy efficiency between UAVs and base stations through UAV trajectory design.

In the context of MEC, due to the limited computational capabilities of UAVs, it is often necessary to offload computing tasks to nearby edge computing devices. Specifically, in time-sensitive scenarios, the paper \cite{ref7} primarily focuses on minimizing service delay and energy consumption. Reference \cite{ref5} considers UAV trajectory optimization for serving Internet of Things (IoT) terminal devices within a finite time.

Literature \cite{ref12} provides a comprehensive introduction to the UAV patrol inspection platform. The application of UAVs in the inspection field offers advantages in efficiency and safety compared to manual methods \cite{ref14}. Recently, several studies have focused on optimizing UAV inspection scenarios. For instance, the authors in \cite{ref13} explore using patrol UAVs as aerial base stations to enhance channel conditions for users located at the cell edges. The paper \cite{ref15} optimizes the three-dimensional trajectories of patrol UAVs passing through fixed points to minimize energy consumption, though it does not consider data processing issues.

Considering the limited onboard energy of UAVs, the work in \cite{ref17} focuses on enhancing the endurance of patrol UAVs through path planning, enabling multiple round-trips to charging stations. The authors in \cite{ref18} propose a deterministic algorithm to minimize patrol time while adhering to battery constraints. Additionally, to efficiently utilize onboard energy, the work in \cite{ref19} studies the maximization of UAV energy efficiency using non-orthogonal multiple access (NOMA) and imperfect CSI.

However, since a single UAV is inefficient in performing large-scale tasks, it is particularly important to jointly perform tasks through multiple UAVs. To address the resource allocation problem of multiple UAVs, the authors in \cite{feng2024graph} propose a novel graph-attention multi-agent trust region reinforcement learning framework for optimizing trajectory design and resource assignment in multi-UAV communication, demonstrating superior convergence and strategy optimization. Literature \cite{ref23} employs an adaptive genetic algorithm for planning UAV swarm missions, aiming to complete tasks using the shortest path and least flight time, though it does not consider energy consumption during this process. To optimize overall UAV energy consumption, the work in \cite{ref21} determines the optimal power allocation strategy through a non-cooperative game model power allocation (NGPA) scheme, aiming to minimize data transmission energy consumption. Furthermore, the authors in \cite{ref22} propose a resource allocation strategy based on the K-means algorithm to enhance overall UAV energy efficiency.

\subsection{Motivations and Main Contributions}
From the above discussion, it can be seen that while many studies have explored improving communication quality or energy efficiency for cellular-connected UAVs by designing their flight trajectories, no literature specifically addresses enhancing overall energy efficiency through balanced task allocation for multiple inspection UAVs performing data collection and offloading tasks simultaneously. Current research on multi-UAV inspections mainly focuses on reducing task completion time and total energy consumption during the inspection process, but overlooks the data offloading aspect. In other words, to ensure the completion of data transmission tasks, UAVs must adjust their flight trajectories to achieve higher communication gains, which in turn lengthens the flight paths and incurs additional energy consumption. Therefore, this paper aims to design an efficient task allocation algorithm that minimizes the energy consumption of UAVs during task execution while ensuring the balanced and efficient completion of data transmission tasks.

Recent research has explored deep learning and generative artificial intelligence (AI) methods, such as multi-task learning and diffusion models, to solve complex optimization problems \cite{yang2020computation,liang2024diffsg}. These methods have demonstrated significant potential in improving efficiency and solution quality for complex network and resource optimization tasks. However, considering the limited computational power of edge devices and the substantial prior information and computing resources these methods often require, we focus on leveraging easily accessible, deterministic information, such as cruise point locations, data offloading volumes, cellular network topology, etc. This allows us to optimize task assignment and UAV trajectories with lower computational time and learning cost while ensuring higher reliability.

In addressing task assignment and energy consumption challenges within the patrol inspection scenario, our approach diverges from existing studies \cite{ref24,ref25,ref26}, presenting a novel balanced task allocation scheme. This paper concentrates on minimizing the total energy consumption and task completion time in a cellular-connected Multiple UAVs-MEC system, with a specific focus on factors such as data size, communication intensity, and the moving distance of multiple UAVs during patrol. The primary objective is to equalize geometry topology, communication rate, and offload volume, thereby enhancing the efficiency and stability of multi-UAV patrols. The key contributions of this paper are outlined as follows:

\begin{enumerate}
\item {This paper proposes a UAV-MEC framework for inspection scenarios, where UAVs capture real-time environmental information and have computing capabilities to make decisions on complex tasks. Considering the high real-time data requirements in patrol scenarios, UAVs must promptly offload collected data to nearby GBSs for processing.}

\item {For the multi-UAV inspection problem, we propose an Energy-Efficient and Balanced Task Assignment Strategy (EBTAS). This strategy uses an improved balanced clustering algorithm to partition the patrol area into sub-regions based on the positions of patrol points, communication conditions, and offload volumes, ensuring an even distribution of tasks. Each sub-region is then treated as a hybrid Time-Energy Traveling Salesman Problem to optimize UAV traversal sequences, minimizing energy consumption and task completion time while ensuring the completion of inspection tasks.}

\item {Numerical results validate the effectiveness of this algorithm, showing advantages in energy consumption and task completion time compared to traditional task allocation strategies.}
\end{enumerate}

\section{System Model and Problem Formulation}

\subsection{System Model}
Consider a cellular-connected UAV inspection system, as illustrated in Fig. 1. This system comprises several patrol UAVs, MEC servers, and data acquisition devices such as high-definition (HD) cameras. We consider a fleet of $N$ (where $N\geq1$) UAVs, denoted by $\mathcal{U}=\left\{u_1, \ldots, u_N\right\}$. A set of $M \geq 1$ GBSs, each equipped with MEC servers, is strategically deployed in the UAV's cruising area and represented by $\mathcal{G}=\left\{g_1, \ldots, g_M\right\}$. The patrol UAV is tasked with traversing $K \geq 1$ cruise points during a flight mission, denoted by $\mathcal{S} = \left\{s_1, s_2, \ldots, s_K\right\}$. The coordinates of cruise points and GBSs are $\mathbf{w}_{s_k} \in \mathbb{R}^{2 \times 1}$ for $s_k \in \mathcal{S}$ and $\mathbf{w}_{g_m} \in \mathbb{R}^{2 \times 1}$ for $g_m \in \mathcal{G}$, respectively.

\begin{figure}[htbp]
  \centering
  \includegraphics[width=0.8\linewidth,keepaspectratio]{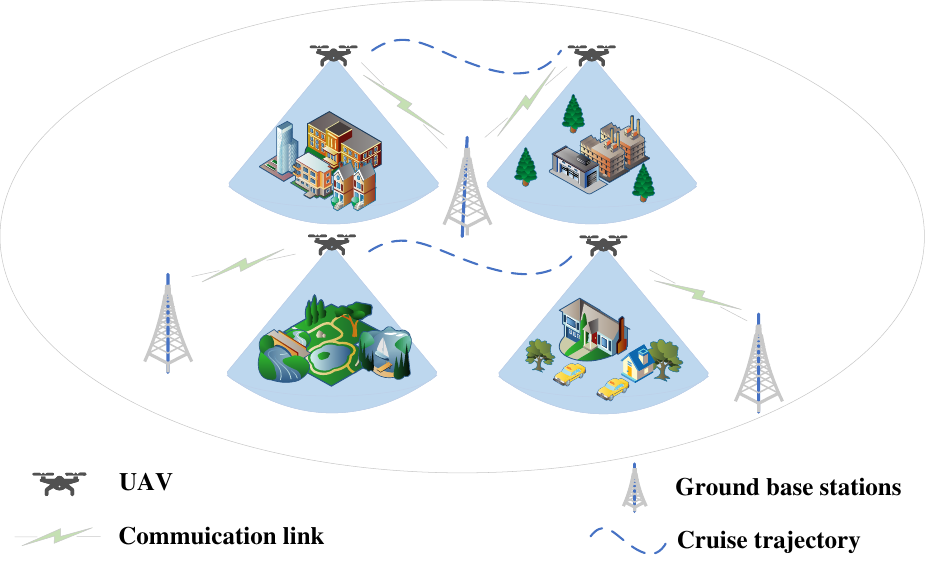}
  \captionsetup{justification=centering,font={small,stretch=0.80},format=hang}
  \caption{Schematic diagram of Multi-UAV Patrol Inspection System}
  \label{fig:center}
\end{figure}

Assuming the existence of a terrestrial network equipped with a sufficient number of GBSs to ensure continuous coverage, we consider a scenario in which a UAV initiates its flight from a predefined starting point $\mathbf{s}_I \in \mathbb{R}^{2 \times 1}$ at an altitude $H_U > 0$. The UAV then navigates through specified cruise points, engaging in data collection and processing tasks. The flight concludes as the UAV lands at the predetermined endpoint $\mathbf{s}_F \in \mathbb{R}^{2 \times 1}$. The sequential path of the UAV's traversal can be represented as $\boldsymbol{\pi} \triangleq {\pi(0), \pi(1), \ldots, \pi(k), \ldots, \pi(K_i), \pi(K_i+1)}$, where $\pi(k) \in \boldsymbol{\pi}$ denotes the index of the $k$-th visited cruise point $s_{\pi(k)} \in \mathcal{S}$ and $K_i$ represents the number of cruise points allocated to the $i$-th UAV. ensuring that $\mathbf{s}_I = s_{\pi(0)} \label{eq:q_I}$, $\mathbf{s}_F = s_{\pi(K_i+1)} \label{eq:q_F}$ and $\sum_{i = 1}^{N}  {K_i = K}$. The overall operational period for the $i$-th UAV is denoted as $T_i$. We can then obtain the average task completion time for all UAVs as $T_{avg}=\frac{1}{N} \sum_{i=1}^{N} T_i$.

During UAV patrol inspections, the UAV methodically visits each designated cruise point, using laser radar or HD cameras to capture crucial information within the target area. This involves acquiring detailed HD photos/videos or 3D point cloud data. Afterward, the UAV proceeds to process this data, aiming to achieve objectives such as target recognition or environment reconstruction relevant to the patrol inspection.

At each cruise point, denoted as $s_k$, the UAV collects $Q_{s_k}$ bits of data. In situations where latency is a critical consideration, it becomes imperative for the UAV to complete the data processing and attain the desired results before reaching the subsequent cruise point. However, the UAV faces constraints attributable to its limited computing capacity and energy resources, which hinder its ability to execute extensive calculations. To overcome this challenge, the strategy of offloading data to GBSs is employed. This approach enables the timely completion of latency-sensitive tasks, ensuring the efficiency and effectiveness of patrol inspection operations.

Without loss of generality, we assume all the GBSs have the same height, denoted as $H_G$. The UAV's position at time $t$ is represented by $\mathbf{\eta}(t)=[x(t), y(t)] \in \mathbb{R}^{2 \times 1}$. Consequently, the distance between the UAV and any given GBS $g_m$ at time $t$ is calculated as
\begin{equation}
d_{g_m}(t)=\sqrt{\left(H_U-H_G\right)^2+\left\|\mathbf{\eta}(t)-\mathbf{w}_{g_m}\right\|^2} .
\end{equation}
The UAV's velocity can be represented as $\mathbf{v}(t) \triangleq \dot{\mathbf{\eta}}(t)$, with the condition $|\mathbf{v}(t)| \leq V_{\max}$ for all $t$ in the interval $[0, T_i]$, where $V_{\max}$ denotes the maximum speed.

The elevation angle between a GBS $g_m$ and the UAV at any given time $t$ is described by the equation:
\begin{equation}
  \theta_{g_m}(t) \triangleq \frac{180}{\pi} \arctan \left(\frac{H_U-H_G}{\left\|\mathbf{\eta}(t)-\mathbf{w}_{g_m}\right\|}\right).
\end{equation}

Based on \cite{ref27}, the real-time communication rate is expressed as:
\begin{equation}
  \begin{aligned}
  R_{g_m}(t)
  & =\left(\chi_3+\frac{\chi_4}{1+e^{-\left(\chi_1+\chi_2 \theta_{g_m}(t)\right)}}\right) \\
  & \times H \log _2\left(1+\frac{\hat{\gamma} P(t)}{\left(d g_m(t)\right)^\alpha}\right),
  \label{eq:R_g_m}
  \end{aligned}
\end{equation}
where $\chi_1<0, \chi_2>0, \chi_4>0$ and $\chi_3+\chi_4=1$. The precise numerical value is contingent upon the characteristics of the propagation environment \cite{ref28}. $\hat{\gamma} = \frac{\beta_0}{\sigma^2 \Lambda}$, where $\beta_0$ represents the average channel power gain at a distance of $1 \mathrm{~m}$, $\sigma^2$ represents the noise power at the base station receiver, and $\Lambda>1$ represents the signal-to-noise ratio gap. $H$ represents the communication bandwidth, $P(t)$ represents the transmission power.

The Time-Division Multiple Access (TDMA) scheme is employed for the patrol UAV to establish connections with the GBSs. We denote $\boldsymbol{\tau}(t)=\left[\tau_{g_1}(t), \ldots, \tau_{g_m}(t), \ldots, \tau_{g_M}(t)\right]$ as an indicator. If $\tau_{g_m}(t)=1$, it indicates that the GBS $g_m$ is scheduled to serve the UAV; otherwise, $\tau_{g_m}(t)=0$.

Define $f_U(t)$ and $f_{g_m}(t)$ as the CPU frequencies at time $t$ for the UAV and the GBS denoted as $g_m$, respectively.
Additionally, the number of CPU cycles required to process per 1-bit of data is denoted as $C_U$, which varies depending on the type of task as indicated in reference \cite{ref30}.

For the GBSs, information causal constraints should be maintained. That is, in any time period $T_P \in[0, T_i]$, the amount of data processed by each GBS is always less than or equal to the amount of data received from the UAV. The constraint expression is as follows:
\begin{equation}
  \int_0^{T_P} \frac{f_{g_m}(t)}{C_U} d t \leq \int_0^{T_P} \tau_{g_m}(t) {R}_{g_m}(t) d t.
  \label{eq:data_constraint}
\end{equation}

The left and right sides of equation (\ref{eq:data_constraint}) represent the total amount of data processed by the GBSs and the amount of data transmitted by the UAV to the GBSs within the time interval $T_P \in [0, T_i]$, respectively.

\subsection{Energy Consumption Model}
The energy consumption of the inspection system is mainly reflected in three aspects: UAV data transmission energy consumption, computing energy consumption, and flight energy consumption. The expression of UAV data transmission energy consumption is given by
\begin{equation}
E_t=\sum_{m=1}^M \sum_{k=1}^{K_i} \int_0^{T_{s_k}} \tau_{g_m}(t) P(t) d t,
\end{equation}
where $\int_0^{T_{s_k}} \tau_{g_m}(t) P(t) d t$ represents the communication energy consumption between the UAV and GBSs $g_m$ during the inspection of cruise point $s_k$.

To optimize the use of limited energy resources, the Dynamic Voltage and Frequency Scaling (DVFS) technique, as cited in reference \cite{ref29}, is implemented in the MEC servers of both the UAVs and the GBSs. 

Drawing from references \cite{ref31} and \cite{ref32}, the computation energy consumption of the UAV is characterized by 
\begin{equation}
E_c=\int_0^{T_i} \vartheta_U f_U^3(t) dt,
\end{equation}
where $\vartheta_U$ denotes the effective capacitance coefficient of the UAV, a parameter intricately linked to the chip architecture of the processor.

Drawing insights from \cite{ref4} and \cite{ref34}, the flight energy consumption model for the rotary-wing UAV is articulated as
\begin{equation}
\begin{aligned}
& E_f=\int_0^{T_i}\left(P_0\left(1+\frac{3\|\mathbf{v}(\mathbf{t})\|^2}{U_{t i p}^2}\right)+\right. \\
& \left.P_i\left(\sqrt{1+\frac{\|\mathbf{v}(t)\|^4}{4 v_0^4}}-\frac{\|\mathbf{v}(t)\|^2}{2 v_0^2}\right)^{\frac{1}{2}}+\frac{1}{2} d_0 \rho s \hat{a}\|\mathbf{v}(t)\|^3\right) d t,
\label{eq:E_f}
\end{aligned}
\end{equation}
where the terms correspond to the blade profile power, induced power, and parasite power.  Specifically, $P_0$ and $P_i$ represent the blade profile power and induced power in a hovering state, respectively. The velocity of the UAV is denoted as $\mathbf{v}(t)$, $U_{tip}$ signifies the tip speed of the rotor blade, and $v_0$ denotes the mean rotor induced velocity during hover.  The variables $d_0$, $\rho$, $s$, and $\hat{a}$ stand for the fuselage drag ratio, air density, rotor solidity, and rotor disc area, respectively.

Consequently, the total energy consumption of the $i$-th UAV is expressed as
\begin{equation}
E_{i}=E_c+E_t+E_f .
\end{equation}

\vspace{-2em}
\subsection{Problem Statement}
For multi-UAV patrol missions, UAVs should capture data and complete computational tasks with minimal energy and time while traversing all patrol points. Moreover, The UAVs should complete their tasks simultaneously in small time intervals to avoid the situation where some UAVs have completed their tasks while others still need a long time to complete their task. Our goal is to minimize the total energy consumption and time during patrol tasks by jointly optimizing the UAV task assignment $\{K_i\}$, traverse order $\{\pi(k)\}$, communication scheduling $\{\tau_{g_m}(t)\}$, task completion time $\{T_i, t_{s_{\pi(k)}}\}$, and UAV trajectory $\{\eta(t)\}$. So the original problem can be expressed as
\[
  (\mathrm{P} 0): \min_{\substack{\{\pi(k)\}\{\mathbf{\eta}(t)\},\\
  \left\{\tau_{g_m}(t)\right\}, t_{s_{\pi(k)}}, T_i, \\
  K_i, \left\{\theta_{g_m}(t)\right\}   }} \sum_{i = 1} ^ {N} \left(E_i + \phi T_i + \lambda\left(T_i - T_{avg}\right)  \right) \\
\]
\begin{subequations}
\allowdisplaybreaks
\begin{align}
   \text { s.t. } 
   &\tau_{g_m}(t) \in\{0,1\}, \sum_{m=1}^M \tau_{g_m}(t) \leq 1, \forall m, t \in[0, T_i], \label{eq:tau}\\
   & \sum_{m=1}^M \int_{t_{s_{\pi(k)}}}^{t_{s_{\pi(k+1)}}} \frac{f_{g_m}(t)}{C_U} d t+\int_{t_{s_{\pi(k)}}}^{t_{s_{\pi(k+1)}}} \frac{f_U(t)}{C_U} d t \geq Q_{s_{\pi(k)}} \nonumber \\
   & ~~~~~~\forall t \in\left[t_{s_{\pi(k)}}, t_{\left.s_{\pi(k+1)}\right]}\right],k=1, \ldots, K_i, \label{eq:qsk}\\
   & \int_0^{T_P} \frac{f_{g_m}(t)}{C_U} d t \leq \int_0^{T_P} \tau_{g_m}(t) {R}_{g_m}(t) d t, \forall m, \\
   & \theta g_m(t)=\frac{180}{\pi} \arctan \left(\frac{H_U-H_G}{\left\|\mathbf{\eta}(t)-\mathbf{w}_{g_m}\right\|}\right), \forall m, \\
   & \|\dot{\eta}(t)\| \leq V_{\max }, \forall t \in[0, T_i], \\
   & \mathbf{\eta}\left(t_{s_{\pi(k)}}\right)=\mathbf{w}_{s_{\pi(k)}}, k=0, \ldots, K_i+1, \label{eq:12f}\\
   & \mathbf{w}_{s_{\pi(0)}}=\mathbf{s}_I, \quad \mathbf{w}_{s_{\pi(K_i+1)}}=\mathbf{s}_F, \label{eq:12g}\\
   & t_{s_{\pi(0)}}=0, \quad t_{s_{\pi(K_i+1)}}=T_i, \label{eq:12h}\\
   &\sum_{i = 1}^{N}  {K_i = K}, \label{eq:12i}
  \end{align}
\end{subequations}
where $\phi$ and $\lambda$ are used as compensation factors to balance the dimensional differences between items. The constraint (\ref{eq:tau}) indicates that each UAV can only communicate with one GBS at a time, while constraint (\ref{eq:qsk}) specifies that the total data processed by the UAV and the GBSs must not be less than the data collected during the inspection. The constraints (\ref{eq:12f}), (\ref{eq:12g}), (\ref{eq:12h}) and (\ref{eq:12i}) ensure that the patrol UAV starts from $\mathbf{s}_I$, traverses all cruise points in the traversal order $\boldsymbol{\pi}$
and finally reaches $\mathbf{s}_F$.

Since the original optimization problem (P0) is not a standard convex programming problem, it is difficult to get the optimal solution directly. In order to solve this problem, we decompose the original optimization problem into two sub-problems: one is the balanced allocation scheme of patrol tasks, and the other is the trajectory optimization between two adjacent cruise points.

\section{Two Sub-problems and Proposed Solutions}
We address the original optimization problem through a two-step approach.  Initially, the tasks assigned to the UAVs are segmented utilizing the proposed energy-efficient and balanced task assignment strategy (EBTAS). Subsequently, in the second step, a technique employing a weighting factor is employed to construct the cruise point traversal sequence, determining the patrol order of the UAV.  The transmission scheduling and trajectory design are then optimized based on the obtained traversal sequence.  Through the iterative execution of these two steps, the original problem can be solved efficiently.

\subsection{Task Assignment}
In the task allocation process for UAVs, our objective is to balance the multiple UAVs' task completion time and minimize the average time while ensuring successful task completion. However, directly calculating the amount of time a drone would take under each task division scenario is quite complex. Considering that the factors affecting the completion time of each UAV mission mainly include the geometry topology, communication rate, and offload volume, we adopt a comprehensive approach by considering the factors mentioned above to achieve a well-balanced task allocation.

In addressing the UAV's movement distance, we account for the actual flight characteristics of UAVs. Typically, UAVs depart from a nearby cruise point and fly to another nearby cruise point to continue their mission. During the task partitioning process, we prioritize the UAV's movement distance from one cruise point to a neighboring cruise point. In practical terms, cruise points that are in close proximity are more likely to be assigned to the same UAV to perform tasks, thereby minimizing energy consumption during the back-and-forth traversal of cruise points.

Therefore, for a cruise point $s_i$, we only need to focus on the position of several nearest cruise points and calculate the average distance of the cruise point $s_i$ from these adjacent cruise points, which can approximate the moving distance of the cruise point in the actual cruise process. However, this method requires computing the distance between pairs of cruise points, which means we have to compute the distance between at least $\frac{1}{2} K (K-1)$ cruise points. Considering the limited computing power of edge computing devices, we adopt a simplified algorithm to calculate the average moving distance of cruise points.

As illustrated in Fig. 2, each cruise point, denoted as \(s_i\), is considered as the origin of coordinates. Within a \(2d \times 2d\) area, we systematically search for the nearest cruise point along the positive and negative directions of the X-axis and Y-axis of the map, adding these identified cruise points to the set of adjacent cruise points. The average distance between the cruise point \(s_i\) and the points in this set is then computed to serve as a reference for determining the actual moving distance of the UAV.

For instance, in Fig. 2, the cruise point \(s_i\) has four red cruise points within the square area as its adjacent set points. Meanwhile, for cruise point \(s_j\), its negative X-axis direction and positive Y-axis direction lead to the same cruise point, resulting in three red cruise points as its adjacent set points, as depicted in the figure.
\begin{figure}[htbp]
  \centering
  \includegraphics[width=0.8\linewidth,keepaspectratio]{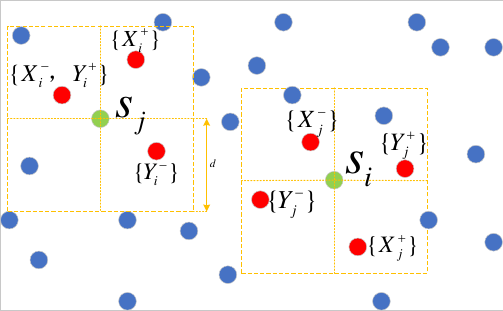}

  \caption{Near cruising point}
  \label{fig:distance}
\end{figure}

For clarity, we introduce the following definitions: the Euclidean distance between cruise point $s_i$ and cruise point $s_j$ is denoted as $\hat d(i,j)$, and the set of neighboring cruise points for cruise point $s_i$ is denoted as $\mathcal{J}(i)=\{X_{i}^{+}, X_{i}^{-}, Y_{i}^{+}, Y_{i}^{-}\}$, where $X_i$ and $Y_i$ denote the vertical and horizontal coordinates of the cruise point $s_i$, $X_i^{+}$ denotes the nearest point in the positive X-axis direction of the cruise point $s_i$, which can be expressed as $X_i^{+}=\min \left\{j \in \mathcal{S} \mid X_j>X_i\right\}$ and similarly for other directions.

Building upon the aforementioned considerations, the distance weight matrix $\mathbf{D}^{k\times1}$ for any given cruise point can be determined. In order to achieve a well-balanced task partitioning, it is essential to minimize both the cumulative movement distance and the discrepancy in movement distances between the UAVs.

In the case of two UAVs, we introduce the 0-1 variable task partition matrices $\mathbf{A}^{k\times1}$ and $\mathbf{B}^{k\times1}$, respectively. In these matrices, the entry $\mathbf{A}(i)$ or $\mathbf{B}(i)$ assumes a value of 1, indicating whether the $i$-th cruise point is assigned to group $\mathbf{A}$ or group $\mathbf{B}$. To illustrate, considering group $\mathbf{A}$ cruise points, the distance of the $i$-th cruise point from nearby points within the same group can be expressed as 
\begin{equation}
  S(i, \mathbf{A})=\sum_{j \in \mathcal{J}(i, \mathbf{A})} \hat d(i, j). 
\end{equation}
A parallel methodology is employed for group $\mathbf{B}$ cruise points.

Formulating the optimization problem, we aim to minimize the combined movement distance and the difference in movement distances between the two UAVs. The optimization problem can be succinctly stated as:
\[
\min_{\mathbf{A}, \mathbf{B}} \sum_{i \in \mathbf{A}} S(i, \mathbf{A}) + \sum_{i \in \mathbf{B}} S(i, \mathbf{B}) + u \left| \sum_{i \in \mathbf{A}} S(i, \mathbf{A}) - \sum_{i \in \mathbf{B}} S(i, \mathbf{B}) \right|
\]
\begin{subequations}
\begin{align}
\text{s.t.} \quad & \mathbf{A}+\mathbf{B}=\mathbf{1}_{k} , \label{eq:2a} \\
& \mathbf{A}, \mathbf{B} \in\{0,1\}^{k} \label{eq:2b}.
\end{align}
\end{subequations}

In the provided equation, the positive parameter $u$ serves as a predefined weight coefficient instrumental for maintaining a balance between the cumulative movement distance and the variance in movement distances among the UAVs.

The constraints \eqref{eq:2a} signify that a given UAV can only be uniquely assigned to one group, thereby promoting a distinct and unambiguous partitioning. On the other hand, constraints \eqref{eq:2b} emphasize that the allocation matrices are comprised solely of 0-1 variables, reinforcing the discrete nature of the assignment variables.

In practical scenarios where cruise points exhibit diverse communication offload volumes attributable to geographical location or task properties, the data process demand on UAVs at each cruise point often differs. To account for this discrepancy, we introduce a weight matrix $\mathbf{Q}^{k\times1}$ for each cruise point. This matrix encapsulates the varying communication offload volume requirements across cruise points.

In an effort to minimize the disparities in communication offload volume allocation between two UAVs, we formulate an optimization problem as follows:
\begin{equation}
  \begin{gathered}
  \min _{\mathbf{A}, \mathbf{B}} \left|\mathbf{A}^T \mathbf{Q}-\mathbf{B}^T \mathbf{Q}\right| \\
  \text{s.t. (4a),(4b).} 
  \end{gathered}
\end{equation}

Moreover, based on the equations (\ref{eq:R_g_m}), we can determine the communication rate of the UAV at any given location with respect to the base station. Therefore, this can establish a communication rate weight matrix for each cruise point, denoted as $\mathbf{R}^{k\times1}$. Similarly, the optimization problem based on the communication weight matrix can be expressed as follows:
\begin{equation}
  \begin{gathered}
  \min _{\mathbf{A}, \mathbf{B}} \left|\mathbf{A}^T \mathbf{R}-\mathbf{B}^T \mathbf{R}\right| \\
  \text{s.t. \eqref{eq:2a},\eqref{eq:2b}.} 
  \end{gathered}
\end{equation}

Currently, we have separately constructed weight matrices based on the UAVs' geometry topology, communication rate and offload volume, along with their respective optimization problems. In order to ensure the stability and balance of the whole system, we integrate the three problems to obtain the following optimization problem:
\[
  (\mathrm{P} 1): \min_{\mathbf{C}} \lVert\mathbf{D}^T\mathbf{C}\rVert _1 + \lVert\mathbf{E}\left(\mathbf{Y}^T\mathbf{C}\mathbf{O}\right)\rVert _1 
\]
\begin{subequations}
\begin{align}
  &\text{s.t.} \quad \mathbf{C}\cdot{1}_2 = {1}_{k}, \label{eq:13a} \\
  & \phantom{\text{s.t.}} \quad\mathbf{C} \in \{0,1\}^{k\times2}. \label{eq:13b}
\end{align}
\end{subequations}
  
In this refined formulation, we present an allocation matrix $\mathbf{C} = (\mathbf{A},\mathbf{B}) \in \{0,1\}^{k\times2}$, where $\mathbf{C}(i, j) = 1$ denotes the assignment of cruising point $s_i$ to UAV $j$.  We denote $\mathbf{E} = \text{diag}(\psi , v, w)$ as a diagonal matrix, where $\psi , v, w$ are predetermined weight coefficients regulating the relative importance of communication rate and offload volume in task allocation.  Let $\mathbf{Y} = (\mathbf{D}, \mathbf{Q}, \mathbf{R})$, and the vector $\mathbf{O} = [1, -1]^T$ is utilized for computing discrepancies.  This methodology not only optimizes task distribution based on geographical positioning but also ensures a balance between the total data offload volume and the communication rate, thereby enhancing the overall network efficiency and stability.

\begin{figure}[!t]
  \centering
  \captionsetup[subfigure]{font=scriptsize} 

  \begin{minipage}[b]{0.3\linewidth}
    \centering
    \subfloat[Cruise point distribution]{\includegraphics[width=1.1\linewidth]{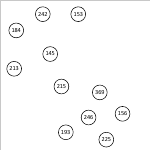}%
    \label{fig:3a}}
  \end{minipage}
  \hfill
  \begin{minipage}[b]{0.3\linewidth}
    \centering
    \subfloat[Clustering result 1]{\includegraphics[width=1.1\linewidth]{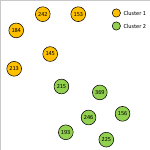}%
    \label{fig:3b}}
  \end{minipage}
  \hfill
  \begin{minipage}[b]{0.3\linewidth}
    \centering
    \subfloat[Clustering result 2]{\includegraphics[width=1.1\linewidth]{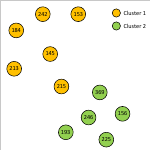}%
    \label{fig:3c}}
  \end{minipage}

  \caption{Diagram of the clustering of cruise points.}
  \label{fig:3}
\end{figure}

However, owing to the inclusion of the $L1$ norm, the aforementioned issue deviates from a standard convex optimization problem, rendering it unsolvable using convex optimization tools such as CVX. Directly addressing it through a brute-force approach, which necessitates computing $2^K$ potential scenarios, imposes a substantial computational burden and proves to be an inefficient solution. In addition, problem (P1) is mainly for the task allocation problem of two UAVs, which cannot be directly extended to the multi-UAV problem. Consequently, we propose an alternative approach based on the principles of the $k$-means algorithm, renowned for its expeditious computation, consistent clustering, and straightforward implementation. At the same time, it has good adaptability to the multi-UAV problem. The general steps of the $k$-means algorithm for classifying a set of $k$ waypoints $\{s_1, s_2, ..., s_k\}$ are as follows:

\begin{enumerate}
    \item The algorithm begins by randomly selecting $n$ samples as initial cluster centers.
    \item Compute the distance from each sample to the cluster centers, using metrics such as Euclidean distance or Manhattan distance. Assign each sample to the cluster corresponding to the nearest cluster center.
    \item Utilize the updated samples in each cluster to recalculate the cluster centers. Let $c_i$ be the cluster center of the $i$-th cluster, and $z$ be the number of samples in the $i$-th cluster.
    \begin{equation}
      c_i=\frac{1}{z} \sum_{j=1}^z s_i.
      \label{eq:c_new}
      \end{equation}
    \item Calculate the sum of squared errors between the cluster centers $\{c_1, c_2, ..., c_n\}$ and the original samples.
    \begin{equation}
      \mathrm{e}=\sum_j^k \sum_i^n\left\|s_j-c_i\right\|^2.
      \end{equation}
    \item If the sum of squared errors reaches a predefined threshold, clustering is complete. Output the cluster centers and the cluster labels corresponding to each sample. Otherwise, return to step (2) and continue iterating until the conditions are met.
\end{enumerate}

To enhance the effectiveness of the k-means clustering algorithm in solving the problem (P1), we introduce several improvements.
Fig. \ref{fig:3} depicts the distribution of cruise point locations and associated communication offload volume. The values within the circles indicate the offload volume requiring computation for each cruise point. When applying the k-means algorithm based on the positions of cruise points, as illustrated in Fig. 3(b), the total offload volume difference between two clustered cruise points is $467$. However, by taking into account the distribution of offload volume and reassigning the cruise point with an offload volume of $215$ in Cluster 2 to Cluster 1, as demonstrated in Fig. 3(c), the resulting offload volume difference between the two clusters is reduced to only $37$.

In order to attain a balanced distribution of offload volume and communication rate among UAVs during the clustering of cruise points using the k-means algorithm, we suggest an enhancement to the distance calculation formula between UAVs and cluster centers.
The feature vector of the cruise point \(s_i\) is defined as \(\mathbf{S}_i=(d_i,q_i,r_i)\), where \(d_i\), \(q_i\), and \(r_i\) are equal to \(\mathbf{w}_{s_i}\), \(\mathbf{Q}(i)\), and \(\mathbf{R}(i)\) respectively. The distance formula between the cruise point and the cluster center is calculated as:
\begin{equation}
L = \kappa_{ij} + m \nu_{ij} + n \varpi_{ij}.
\label{eq:distance}
\end{equation}
where \(m\) and \(n\) are the weight coefficients, respectively, and \(\kappa_{ij}\), \(\nu_{ij}\), and \(\varpi_{ij}\) represent the average variance of all clusters after the cruise point \(s_i\) is added to cluster \(j\).
\begin{equation}
  \left\{
  \begin{aligned}
      \kappa_{ij} &= \frac{1}{z_j} \sum_{i=1}^{z_j} \left(d_i - \bar{d}\right)^2 \\
      \nu_{ij} &= \frac{1}{z_j} \sum_{i=1}^{z_j} \left(q_i - \bar{q}\right)^2 \\
      \varpi_{ij} &= \frac{1}{z_j} \sum_{i=1}^{z_j} \left(r_i - \bar{r}\right)^2,
  \end{aligned}
  \right.
  \end{equation}
  where \(z_j\) is the number of cruise points in the $j$-th cluster, and \(\bar{d}, \bar{q}, \bar{r}\) are the mean values of all cruise points in the cluster.

Moreover, given the necessity for a comprehensive evaluation of cruise point clustering, which takes into account factors like the geometry topology, communication rate, and offload volume, relying solely on the calculation of the sum of squared errors is insufficient for determining appropriate clustering that adequately reflects task allocation balance. Therefore, we use the objective function in problem (P1) as the formula for calculating the deviation between clusters.

Additionally, due to the limited energy resources of UAVs, namely, the constrained quantity of cruise points per cluster, we introduce the concept of a maximum allowable number of cruise points per cluster, denoted as $c_{max}$, for post-processing the outcomes of the clustering algorithm. In the case of clusters exceeding the specified limit $c_{max}$, assuming an excess number denoted as \(\hat n\), the \(\hat n\) cruise points farthest from the cluster center are excluded, and reallocated to the nearest cluster with a sample size below $c_{max}$. This ensures that the sample size of each cluster remains within acceptable limits, simultaneously minimizing the impact on the original clustering results. The refined algorithmic process is presented in Algorithm 1.

\begin{algorithm}[h]
  \caption{Cruise Point Clustering}\label{alg:uav_clustering}
  \begin{algorithmic}[1]
    \REQUIRE State information $\mathbf{S}_i$ of $K$ cruise points, number of clusters $n$, maximum sample size per cluster $c_{\text{max}}$, maximum iterations $max\_iter$
    \ENSURE Clustering result $result$, cluster centers $c$
    \STATE Randomly select $n$ cruise points as initial cluster centers $c=\left\{c_1, c_2, \ldots, c_n\right\}$
    \STATE Initialize $c_{\text{new}}=\{\}$ and $iter=0$
    \WHILE{true} \label{line:while_start}
      \FOR{$i=1$ to $K$}
        \STATE Compute distances between cruise point $i$ and cluster centers using Eq. (\ref{eq:distance})
        \STATE Assign cruise point $i$ to the nearest cluster center $c_j$
        \STATE Set $result(i, j) = 1$
      \ENDFOR
      \FOR{$j=1$ to $n$}
        \STATE Update cluster center $c_{\text{new}}$ using Eq. (\ref{eq:c_new})
        \IF{$z_j > c_{\text{max}}$}
          \STATE Select $z_j - c_{\text{max}}$ farthest points from cluster $j$
          \STATE Reassign these points to the nearest clusters with sample size $< c_{\text{max}}$
        \ENDIF
      \ENDFOR
      \STATE Compute average deviation $p$ among all pairs of clusters using the  objective function of (P1) \label{line:calculate_deviation}
      \IF{$p < p_{\text{threshold}}$ \textbf{or} $iter \ge max\_iter$}
        \STATE \textbf{Break}
      \ENDIF
      \STATE $iter \gets iter + 1$
    \ENDWHILE
  \end{algorithmic}
\end{algorithm}

Algorithm 1 is an improved version of the traditional k-means clustering algorithm, offering relatively low time complexity. Specifically, the overall time complexity is $\mathcal{O}(max\_iter \cdot n^2 \cdot K)$. In most practical applications, the parameters $max\_iter$ (maximum iterations), \(n\) (number of patrol UAVs),  and  \(K\) (number of cruise points) are typically within a controllable range, ensuring that the algorithm remains efficient.

\subsection{Path planning}
The majority of the UAV's trajectory is determined by the order of accessing cruise points, and thus, the traversal sequence to some extent influences the energy consumption of the UAV. Additionally, during the flight, the UAV offloads the collected data to a GBS for processing. The traversal sequence affects the communication rate between the UAVs and GBSs, thereby influencing the data offloading time, which is also closely related to the data size. In this chapter, a novel approach is taken to design the visit sequence of the UAV \cite{ref27}, which takes into account the flight distance of the UAV and the topology between the cruise point and the GBSs.

Initially, a directed weighted graph $\boldsymbol{\Omega} \triangleq \{V, F\}$ is constructed based on the topological structure of cruise points and the GBSs. The vertex set $V$ is represented as:
\begin{equation}
V=\{s_0, \ldots, s_k, \ldots, s_{K_i}, s_{K_i+1}\},
\end{equation}
where $s_0$ denotes the starting point, and $s_{K_i+1}$ represents the endpoint. The edge set $F$ is represented as:
\begin{equation}
F=\{(s_i, s_j), i \neq j; i, j \in \mathcal{K}\}.
\end{equation}

Considering both the distance and speed, the weight factor for edges represented as:
\begin{equation}
W_{s_i, s_j}=Q_{s_i} \cdot \frac{E_{s_i, s_j}}{Q_{s_i, s_j}}, \quad \forall i, j \in \mathcal{K}, i \neq j,
\end{equation}
where $E_{s_i, s_j}$ represents the energy consumption of the UAV when flying along the edge $\left(s_i, s_j\right)$. $Q_{s_i, s_j}$ represents the actual communication offload volume when the UAV flies along the edge $\left(s_i, s_j\right)$ and $Q_{s_i}$ represents the captured data volume at cruise point $s_i$.
Thus, the equation for this problem is obtained:
\[
  (\mathrm{P} 2): \min \sum_{i \in \mathcal{K}} \sum_{j \in \mathcal{K}, i \neq j} \xi_{i, j} W\left(s_i, s_j\right) \\
\]
\begin{subequations}
\begin{align}
  \text{s.t.} 
  & \xi_{i, j} \in\{0,1\}, \forall i, j \in \mathcal{K}, i \neq j, \label{eq:20c}\\
  & \xi_{i, j}+\xi_{j, i} \leq 1, \forall i, j \in \mathcal{K}, i \neq j, \label{eq:20d}
\end{align}
\end{subequations}
where $\xi_{i, j}$ is a binary variable representing the selection of the path. If $\xi_{i, j}=1$, it indicates that the UAV flies from $i$ along the edge $(i, j)$ to $j$, otherwise, this path is not chosen.

After delineating the task scope and the corresponding traversal order of the UAV, the specific path planning for the UAV can be decomposed into $K_i$+ 1 independent subproblems, specifically addressing the optimal path between two consecutive cruise points.

The objective of the patrol inspection mission is to traverse all cruise points, capture data, and complete computational tasks within a reasonable time while minimizing energy consumption.
Building upon the previously introduced UAV energy consumption model, we proceed to conduct joint optimization involving the duration $T_{s_{\pi(k)}}^{s_{\pi(k+1)}}$, the trajectory of the patrol drone denoted as $\mathbf{\eta}(\mathbf{t})$, CPU frequencies $f_{g_m}(t)$ and $f_U(t)$, along with the scheduling factor $\tau_{g_m}(t)$. The optimization problem can be formulated as:

\begin{equation}
  \begin{aligned}
   (\mathrm{P} 3): &\min _{\substack{\{\mathbf{\eta}(t)\},\left\{\tau_{g_m}(t)\right\}, T_{s_\pi(k)}^{s_\pi(k+1)},\left\{\theta_{g_m}(t)\right\}, \\
  \left\{f_U(t), f_{g_m}(t)\right\}}} E_i\left[t_{s_{\pi(k)}}, t_{s_{\pi(k+1)}}\right] \\
     \text { s.t. } 
        &\sum_{m=1}^M \int_0^{T_{s_{\pi(k)}}^{s_{\pi(k+1)}}} \frac{f_{g_m}(t)}{C_U} d t+\int_0^{T_{s_{\pi(k)}}^{s_{\pi(k+1)}}} \frac{f_U(t)}{C_U} d t  \\
        &~~~~~~~~~~~~~~~~~~~~~\geq Q_{s_{\pi(k)}},\forall t \in\left[0, T_{s_{\pi(k)}}^{s_{\pi(k+1)}}\right], 
        \label{eq:P2}
  \end{aligned}
\end{equation}
where the constraint specifies that the cumulative volume of data collaboratively processed by the patrol UAV and GBSs must exceed the amount of data captured by the UAV at cruise point $s_{\pi(k)}$ during the flight between two consecutive cruise points $s_{\pi(k)}$ and $s_{\pi(k+1)}$.
By introducing the path discretization technique, Sequential Convex Approximation (SCA) technique, and Block Coordinate Descent (BCD) framework, this problem can be transformed into a standard convex optimization problem, which can be solved by convex optimization solver such as CVX \cite{ref27}.

\color{black}
\section{Simulation Results}
In this section, we assess the effectiveness of the proposed energy-efficient and balanced task assignment strategy (EBTAS) in a patrol inspection scenario involving several UAVs. The related simulation parameters are summarized in Table \ref{tab:parameters}.

\begin{table}[!t]
  \caption{Related Simulation Parameters\label{tab:table1}}
  \centering
  \label{tab:parameters}
  \begin{tabular}{|c|c|c|}
  \hline
  \textbf{Notation} & \textbf{Definition} & \textbf{Value} \\
  \hline
  $H$ & Channel bandwidth (MHz) & 1 \\
  \hline
  $\beta_0$ & Average channel power gain at $d_0=1 \mathrm{~m}$ (dB) & -50 \\
  \hline
  $\sigma^2$ & Noise power at the receiver (dBm) & -100 \\
  \hline
  $C_{U}$ & \begin{tabular}[c]{@{}l@{}}The required number of CPU \\ cycles per bit (cycles/bit)\end{tabular} & $1000$ \cite{9090334} \\
  \hline
  $\vartheta_U$ & Effective capacitance coefficient of the UAV & $10^{-27}$ \cite{8873672} \\
  \hline
  $P(t)$ & UAV transmitting power (W) & 0.1 \\
  \hline
  $P_0$ & Blade profile power in hovering status (W) & 79.8 \cite{9311644} \\
  \hline
  $P_i$ & Induced power in hovering status (W) & 88.6 \\
  \hline
  $v_0$ & Mean rotor induced velocity in hover (m/s) & 4 \\
  \hline
  $d_0$ & Fuselage drag ratio & 0.6 \\
  \hline
  $\rho$ & Air density (kg/$\mathrm{m}^3$) & 1.2 \\
  \hline
  $s$ & Rotor solidity & 0.05 \\
  \hline
  $\hat{a}$ & Rotor disc area ($\mathrm{m}^2$) & 0.5 \\
  \hline
  $U_{\text{tip}}$ & Tip speed of the rotor blade (m/s) & 120 \cite{9311644} \\
  \hline
  $H_U$ & Altitude of UAV (m) & 25 \cite{9149835} \\
  \hline
  $H_G$ & Altitude of GBS (m) & 100 \cite{9149835} \\
  \hline
  $f_U(t)$ & UAV CPU frequency (GHz) & 0.8 \\
  \hline
  $f_{g_m}(t)$ & GBS CPU frequency (GHz) & 8 \\
  \hline
  \end{tabular}
\end{table}

We first conducted a simulation involving task partitioning with 20 cruise points ($k=20$) and two UAVs. In practical multi-UAV patrol scenarios, task allocation typically follows either geographical area delineation \cite{ma2023multi} or the minimization of the UAVs' shortest moving distances \cite{rovira2017predictive}. When employing the geographical area criterion, each UAV exclusively undertakes tasks within a predefined area, neglecting considerations such as communication offload volume requirements of cruise points within the mission range and communication rate. Alternatively, the criterion of minimizing the UAVs' shortest moving distances aims to reduce flight energy consumption, albeit without accounting for the energy consumption associated with data transmission and processing at cruise points within the mission range. This paper establishes a comparative analysis between these two task partitioning strategies as a baseline.

\begin{figure}[!t]
\centering
\includegraphics[width=3in]{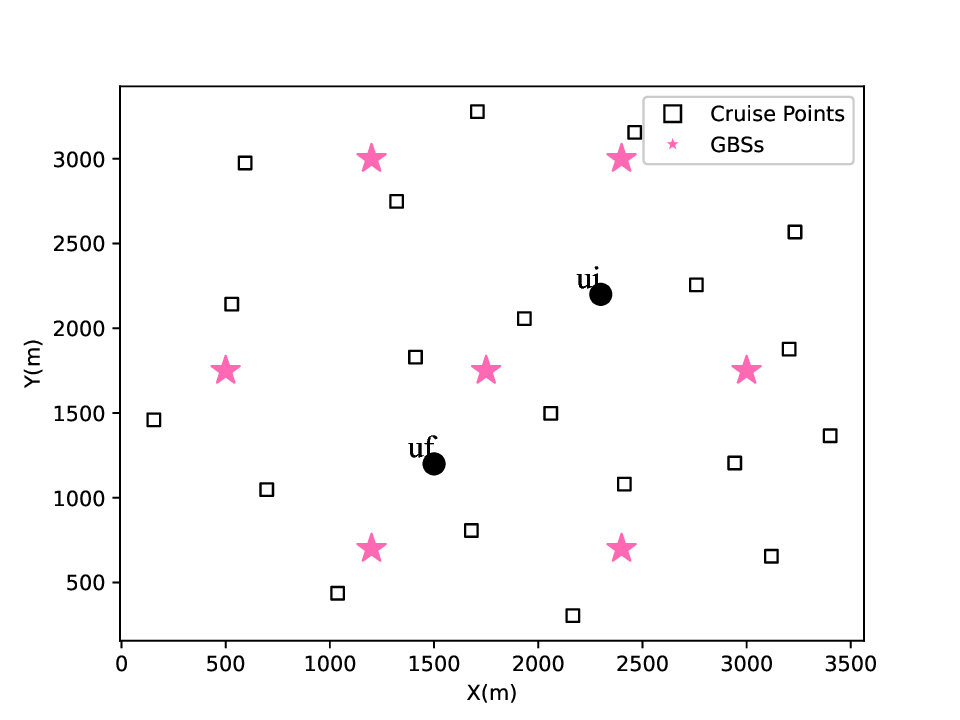}
\caption{Distribution of cruise points and GBSs.}
\label{fenbu}
\end{figure}

Fig. \ref{fenbu} illustrates a simulation of a realistic patrol scenario, wherein cruise points are approximately uniformly distributed across the entire map, and GBSs exhibit a honeycomb-like pattern. To emulate practical conditions, specific starting and ending points for the UAVs are designated, ensuring that each UAV initiates and concludes its mission at the same specified points.

For graphical representation, the starting points ($u_i$) and ending points ($u_f$) are denoted by black solid circles, the cruise points are represented by rectangular boxes, and the GBSs are depicted as pink pentagrams.

We initially contemplate the UAV flight trajectories under three task allocation strategies, each characterized by an equivalent communication offload volume at all cruise points. In the scenario where $Q_{s_k}$=100Mbits, as depicted in Fig. 5(a), task allocation is conducted based on geographical regions. Specifically, UAV $a$ traverses and cruises the upper half of the map, while UAV $b$ performs similar activities in the lower half. Due to the uncertainty in the distribution of cruise points, the overall travel distance is relatively extended. In Fig. 5(b), UAV task allocation is predicated on the shortest moving path, resulting in a reduced overall travel distance during the cruise point traversal, thereby minimizing flight energy consumption. 

\begin{figure*}[!t]
  \centering
  \captionsetup[subfloat]{font=scriptsize} 
  \subfloat[Region criterion]{\includegraphics[width=2.3in]{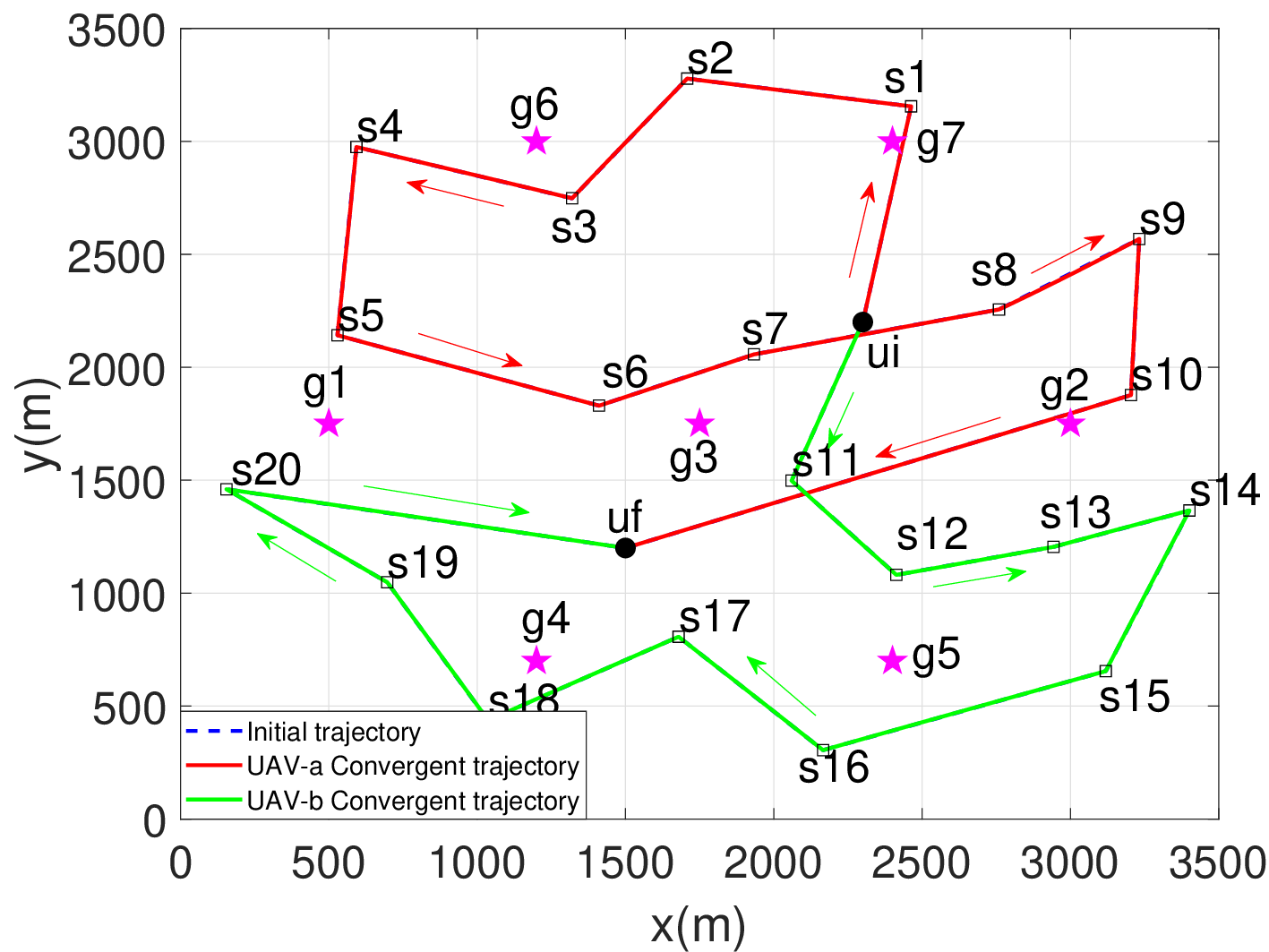}%
  \label{fig:left}}
  \hfil
  \subfloat[Shortest distance criterion]{\includegraphics[width=2.3in]{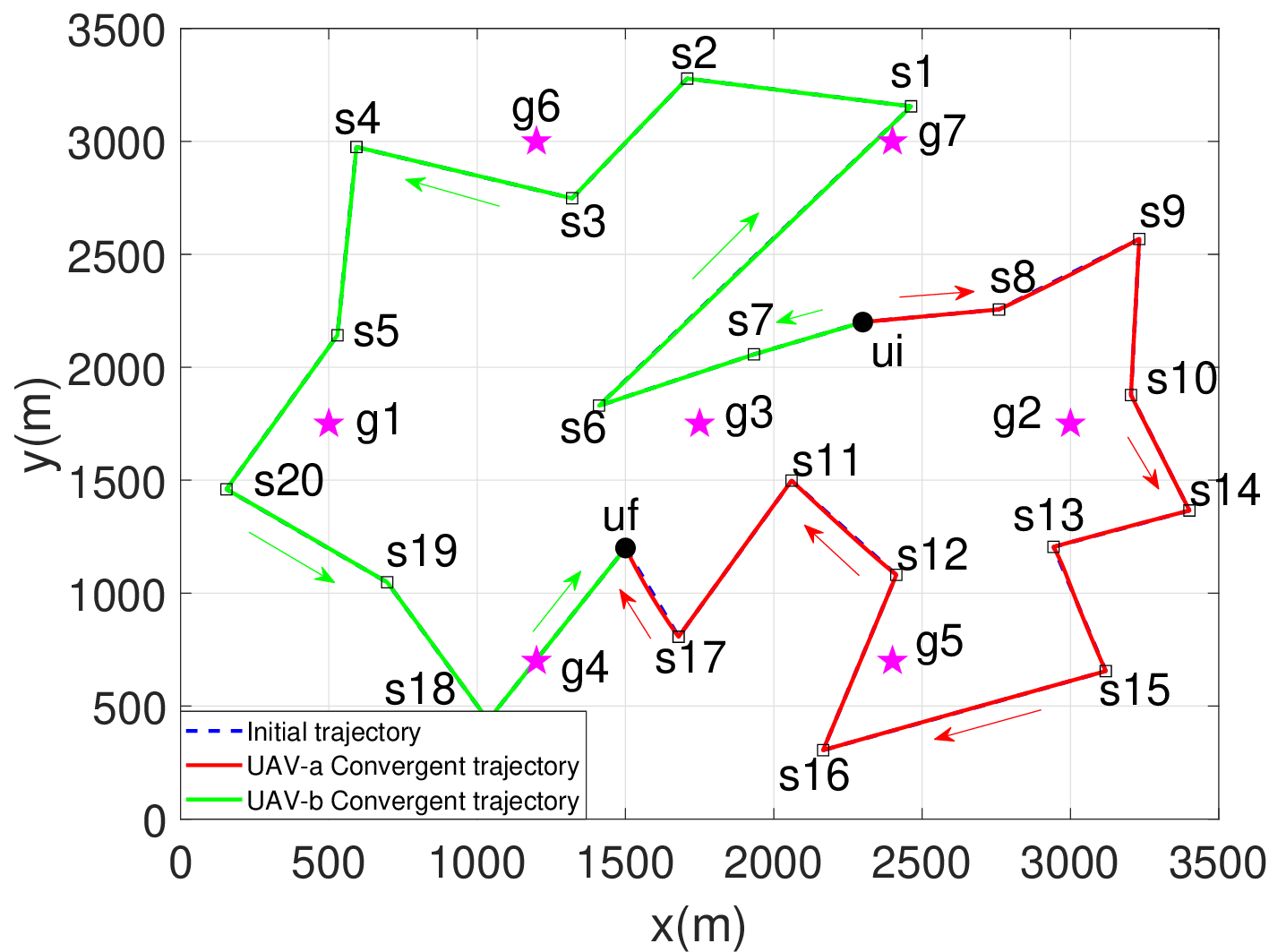}%
  \label{fig:center}}
  \hfil
  \subfloat[EBTAS]{\includegraphics[width=2.3in]{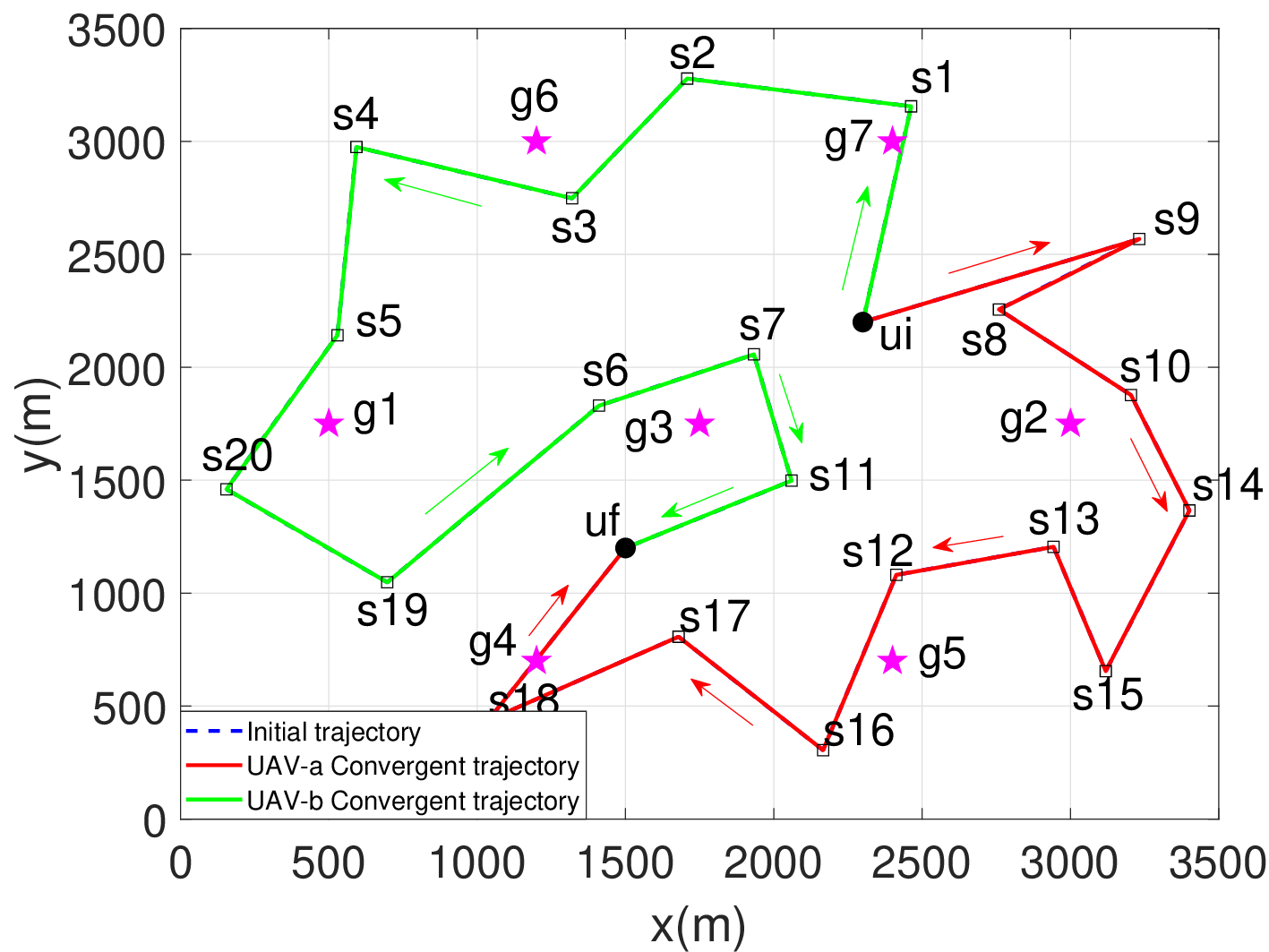}%
  \label{fig:right}}

  \vspace{1em}
  \footnotesize{$Q_{s_k}$=100 Mbits.}
  \caption{Optimized route of cruise points with two UAVs.}
  \label{fig:both}
\end{figure*}

\begin{figure}[!t]
  \centering
  \captionsetup[subfloat]{font=scriptsize} 
  
  \subfloat[Region criterion]{\includegraphics[width=1.5in]{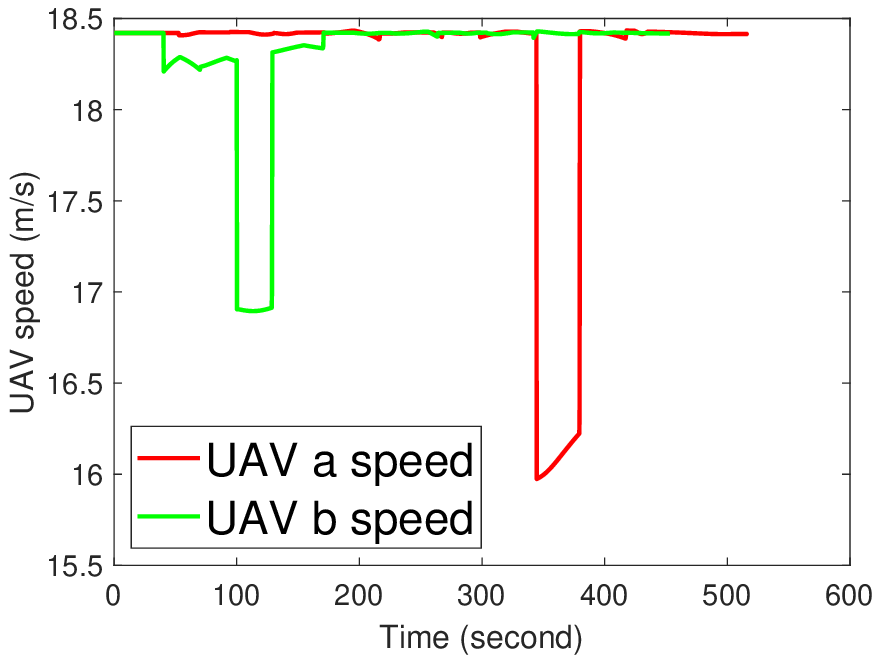}%
  \label{fig:5a}}
  \hfil
  \subfloat[Region criterion]{\includegraphics[width=1.5in]{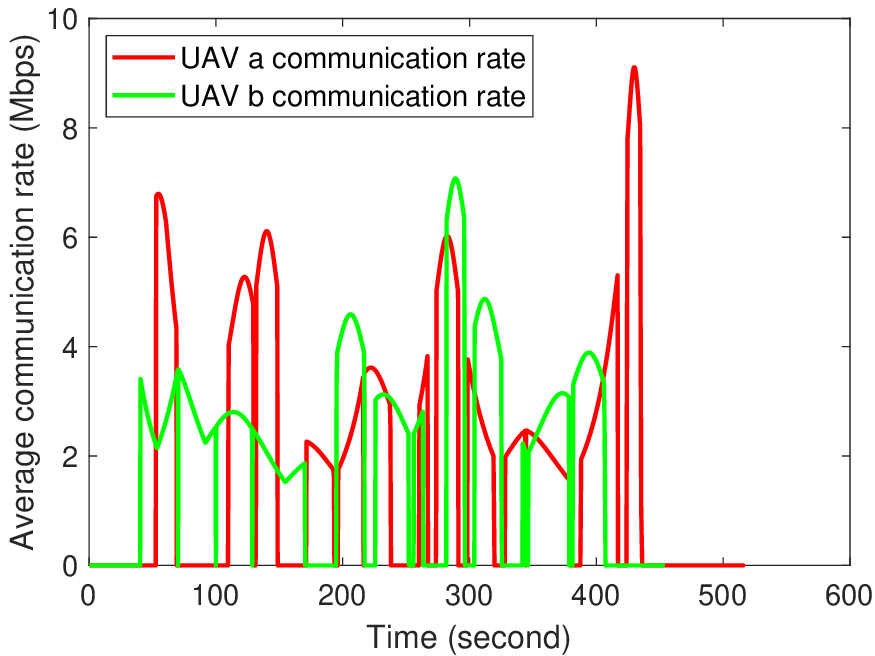}%
  \label{fig:5d}}
  \hfil
  
  \vspace{-1em}

  \subfloat[Shortest distance criterion]{\includegraphics[width=1.5in]{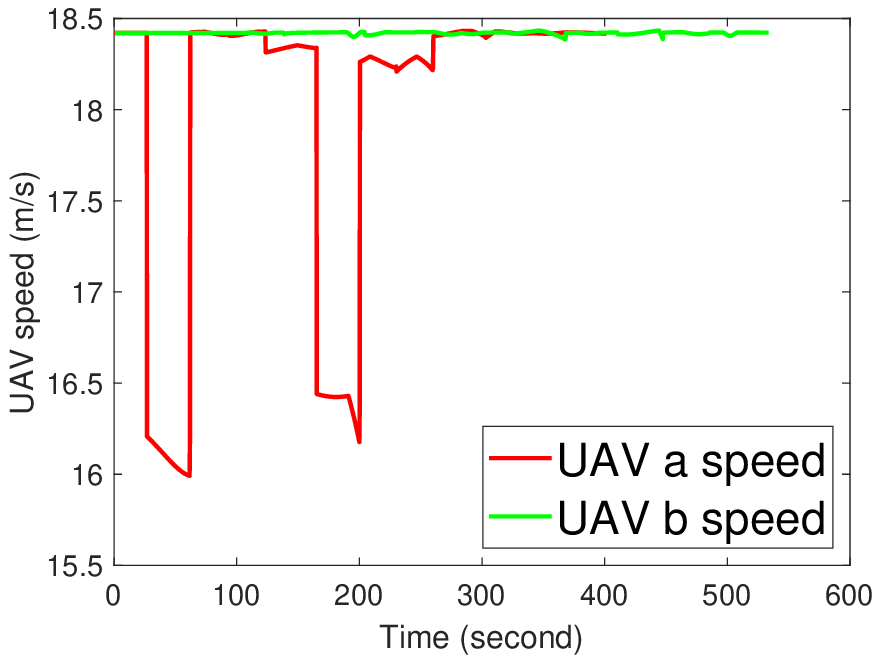}%
  \label{fig:5b}}
  \hfil
  \subfloat[Shortest distance criterion]{\includegraphics[width=1.5in]{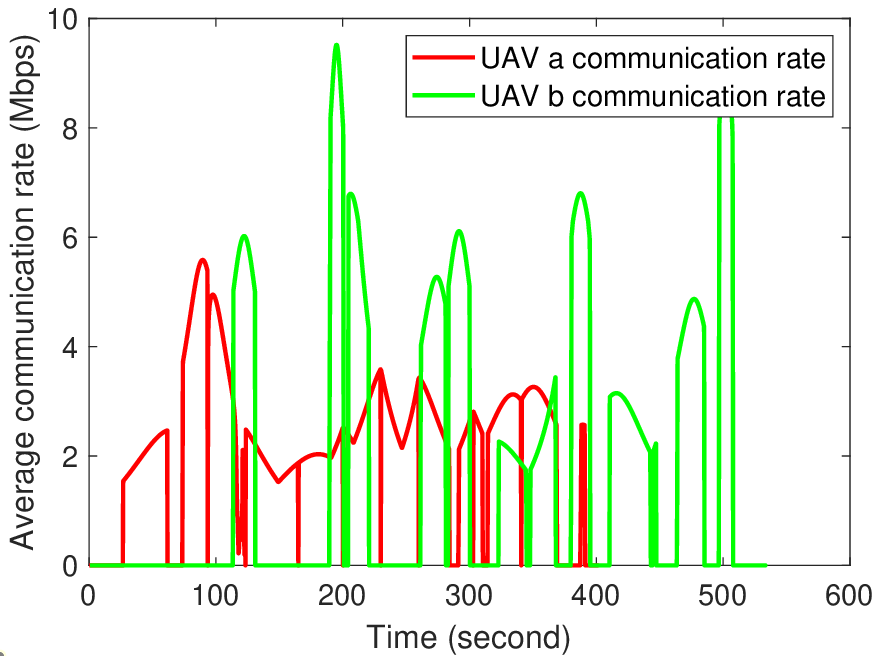}%
  \label{fig:5e}}
  \hfil
  
  \vspace{-1em}
  \subfloat[EBTAS]{\includegraphics[width=1.5in]{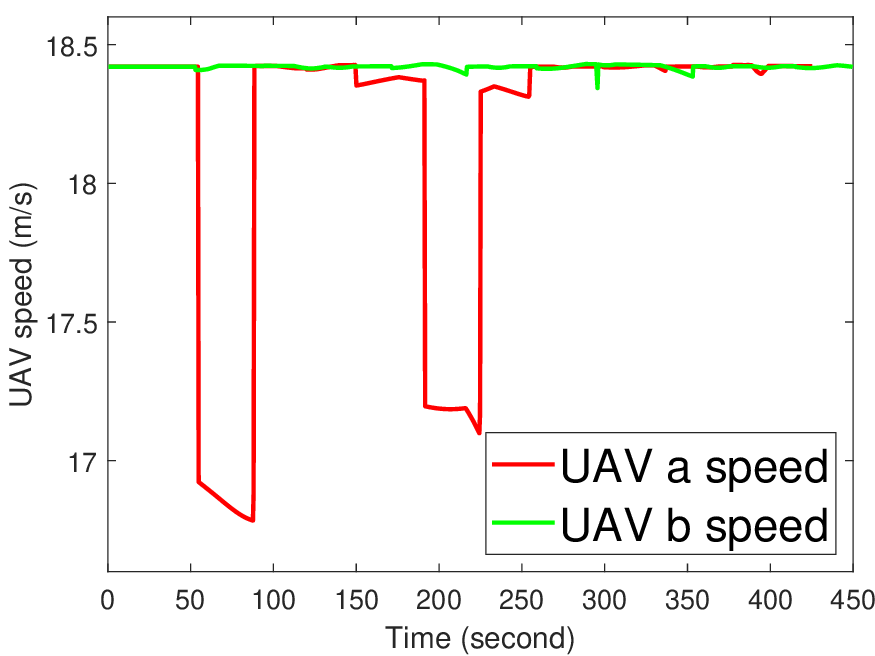}%
  \label{fig:5c}}
  \hfil
  \subfloat[EBTAS]{\includegraphics[width=1.5in]{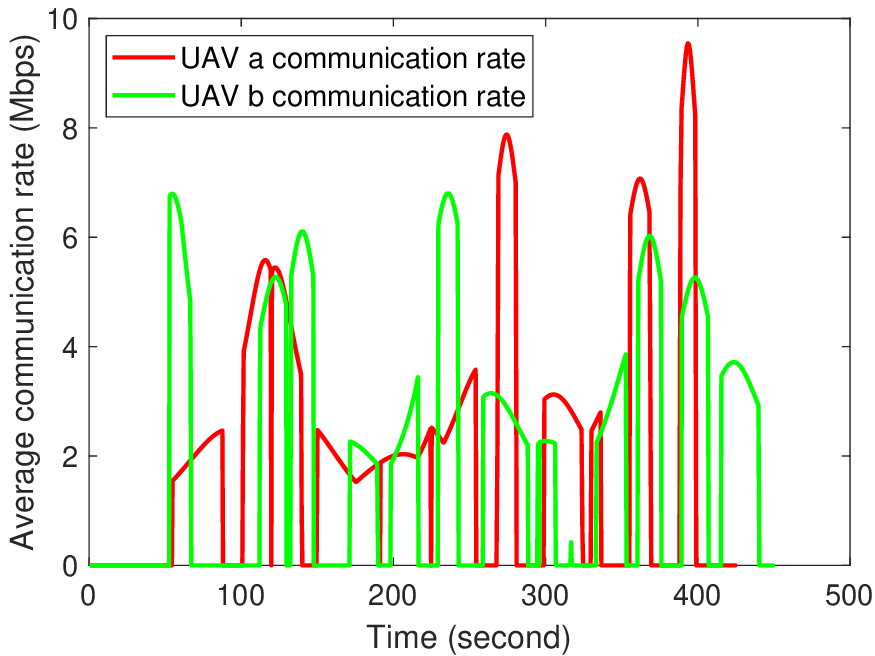}%
  \label{fig:5f}}

  \footnotesize{$Q_{s_k}$=100 Mbits.}
  
  \captionsetup{justification=centering} 
  \caption{UAVs' speed and communication rate over time.}
  \label{fig:8}
\end{figure}

As illustrated in Fig. 5(c), UAVs are partitioned based on the EBTAS proposed in this study. It is observed that the division and traversal sequence of certain cruise points, such as $s_{8}-s_{10}$, $s_{7}-s_{11}$,  differ from those in Fig. 5(b). These particular road segments are closer to the GBSs, resulting in a stronger communication gain, which can effectively reduce the energy consumption associated with UAV computation and data transmission. Furthermore, owing to the insufficient communication offload volume, the space for trajectory optimization is limited, leading to predominantly linear flight paths along most road segments.

\begin{figure}[!t]
  \centering
  \captionsetup[subfigure]{font=scriptsize} 
  \begin{minipage}[b]{0.9\linewidth}
    \centering
    \captionsetup{justification=centering}
    \subfloat[Total system energy consumption]{\includegraphics[width=0.8\linewidth]{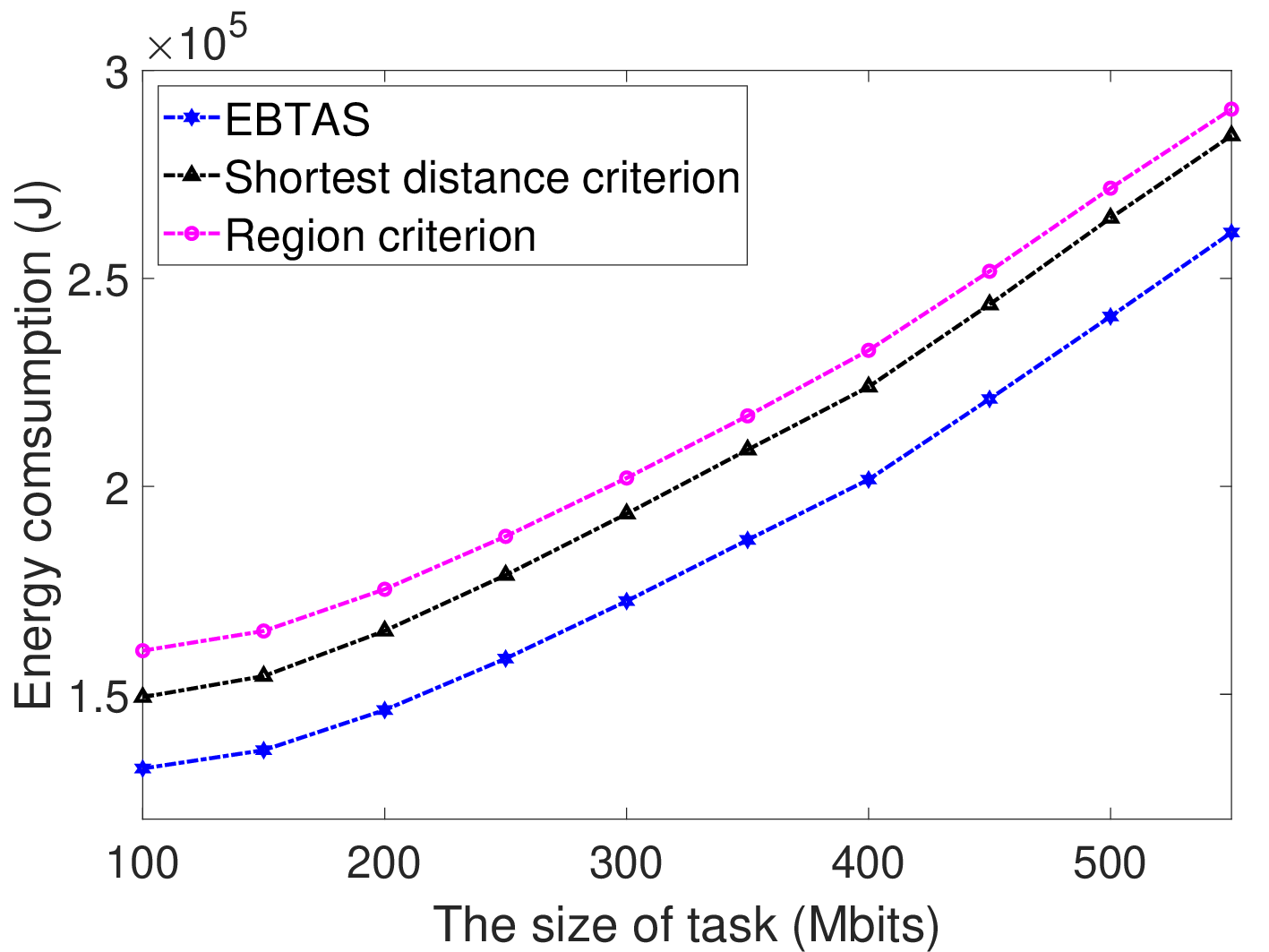}%
    \label{fig:6a}}
  \end{minipage}
  \hfill
  \\
  \begin{minipage}[b]{0.9\linewidth}
    \centering
    \subfloat[Mission completion time]{\includegraphics[width=0.8\linewidth]{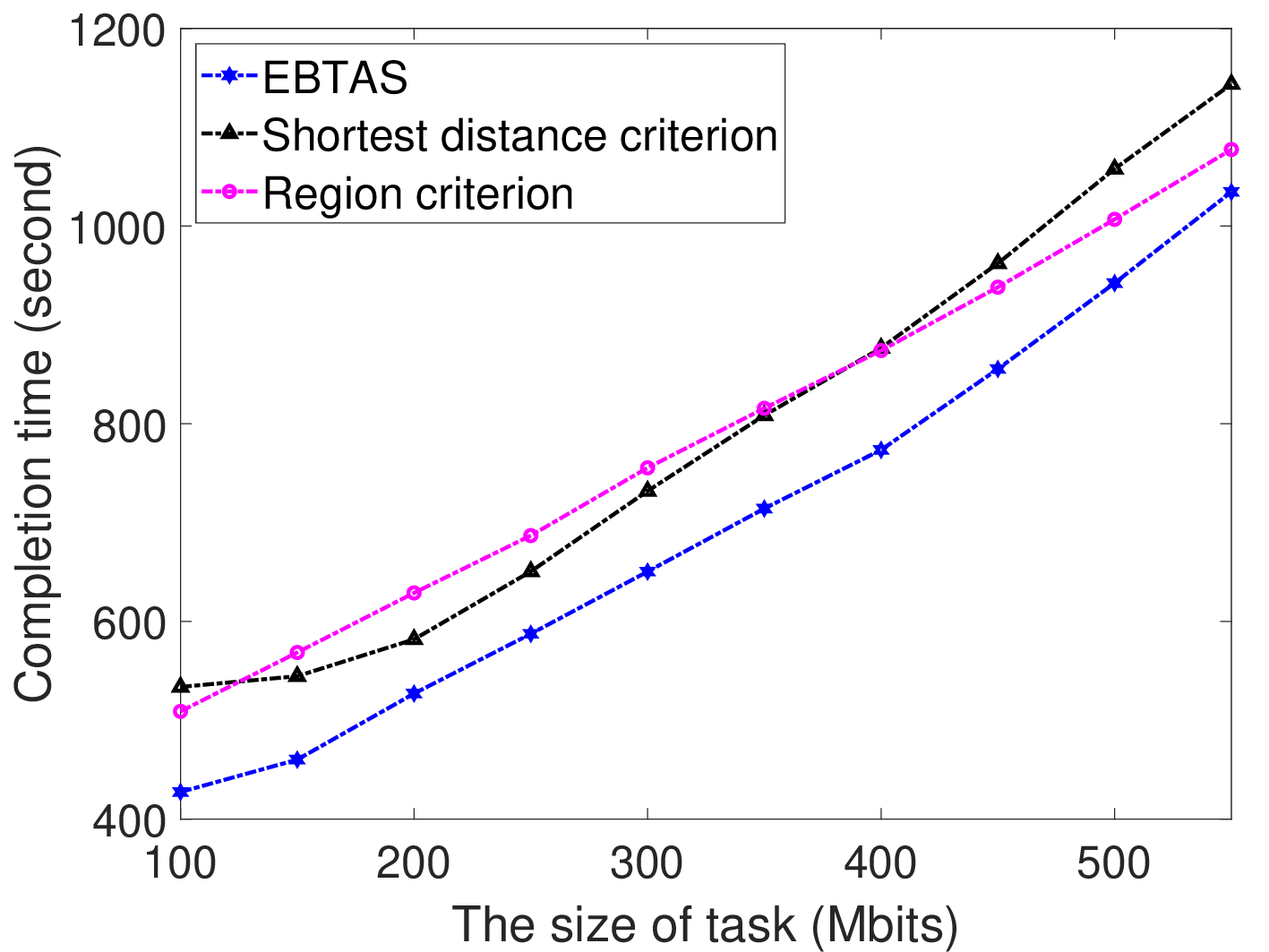}%
    \label{fig:6b}}
  \end{minipage}

  \captionsetup{justification=centering}
  \caption{Energy consumption and mission completion time versus task size $Q_{s_k}$.}
  \label{fig:6}
\end{figure}

Additionally, Fig. \ref{fig:8} illustrates the correlation between the task completion time of the UAV and its speed, as well as the communication rate. The analysis reveals that when the data offloading at each cruise point is minimal, the UAV predominantly operates at maximum speed along most sections of the route, as depicted in Fig. \ref{fig:8}(a), Fig. \ref{fig:8}(c), and Fig. \ref{fig:8}(e).  Meanwhile, Fig. \ref{fig:8}(b), Fig. \ref{fig:8}(d), and Fig. \ref{fig:8}(f) show that there is a communication interruption during the flight of the UAV, indicating that the UAV does not offload data during this period. This occurrence is attributed to the considerable distance between the UAV and the GBS, rendering data offloading unfeasible. In other words, when the computational task is relatively small, the UAV autonomously executes the task, resulting in greater energy efficiency. However, when confronted with a substantial computational load, the UAV adjusts its trajectory towards the GBSs to ensure a stable communication connection.

In the scenario where the cruise points have the same communication offload volume, Fig. \ref{fig:6} shows the energy and time consumption of each UAV under the three assignment strategies. The numerical results show that the proposed EBTAS is superior to the traditional assignment strategy in terms of task completion time and energy consumption.

\begin{figure*}[!t]
  \centering
  \captionsetup[subfloat]{font=scriptsize} 
  \subfloat[Region criterion]{\includegraphics[width=2.3in]{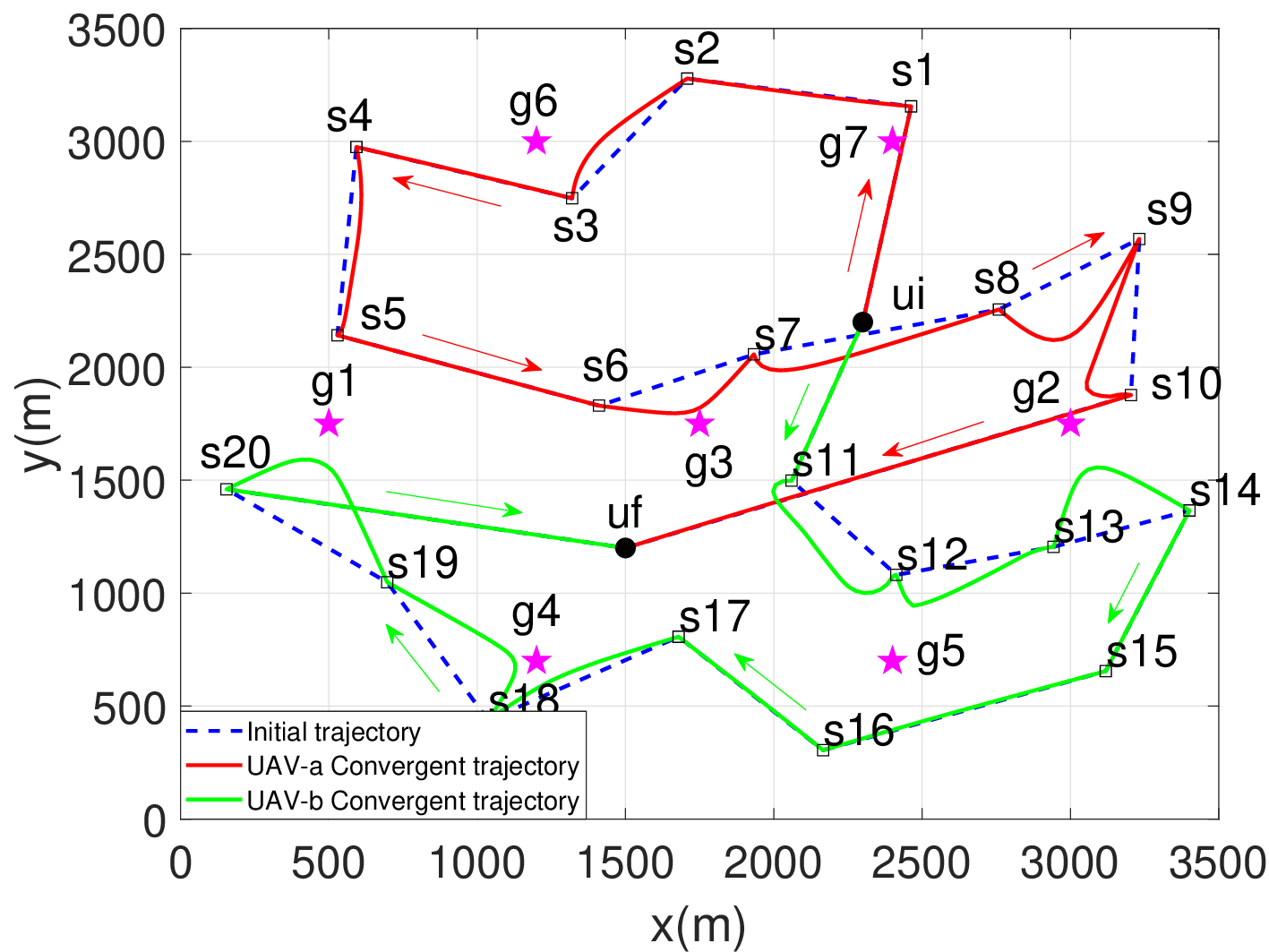}%
  \label{fig:left}}
  \hfil
  \subfloat[Shortest distance criterion]{\includegraphics[width=2.3in]{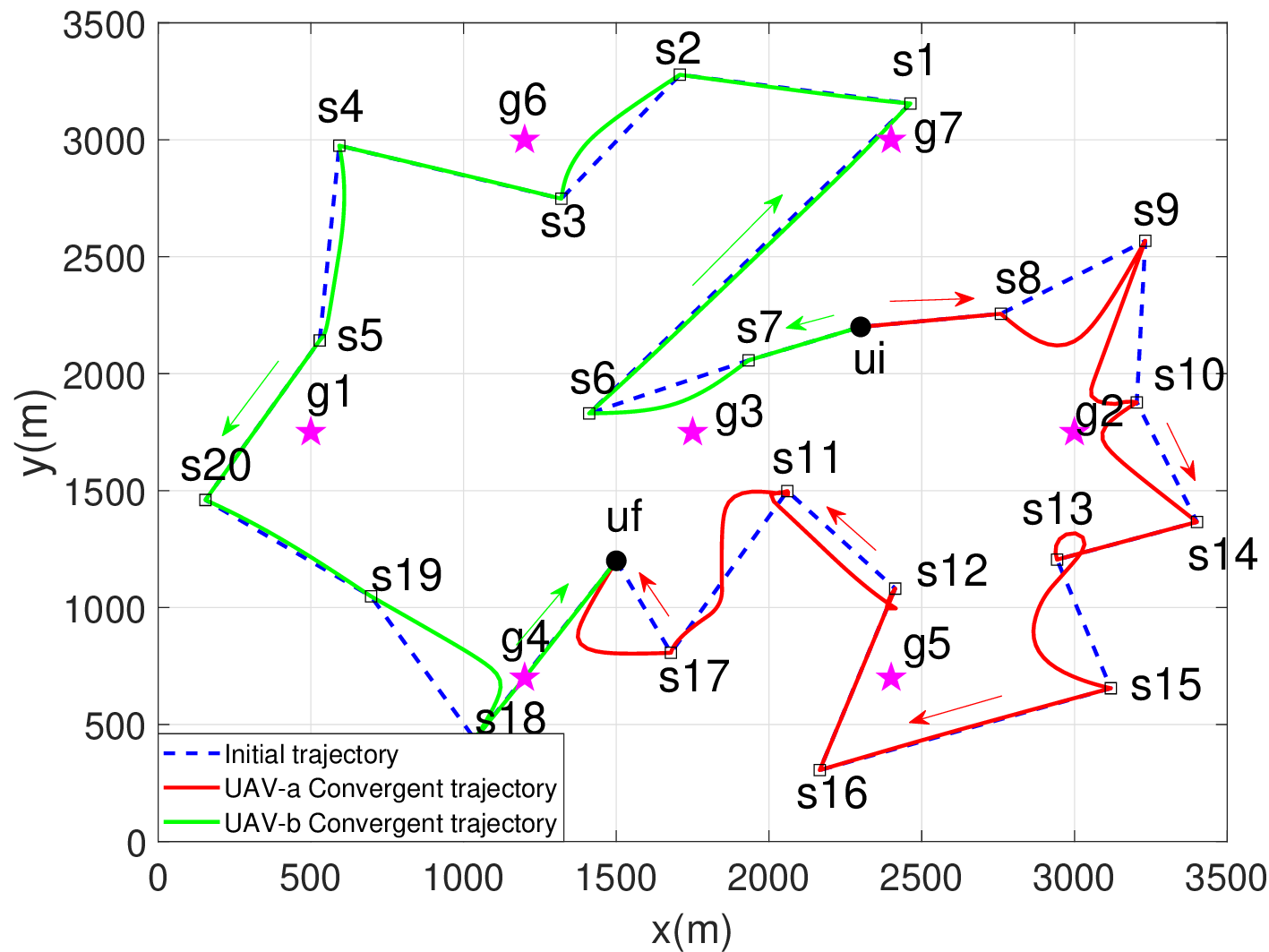}%
  \label{fig:center}}
  \hfil
  \subfloat[EBTAS]{\includegraphics[width=2.3in]{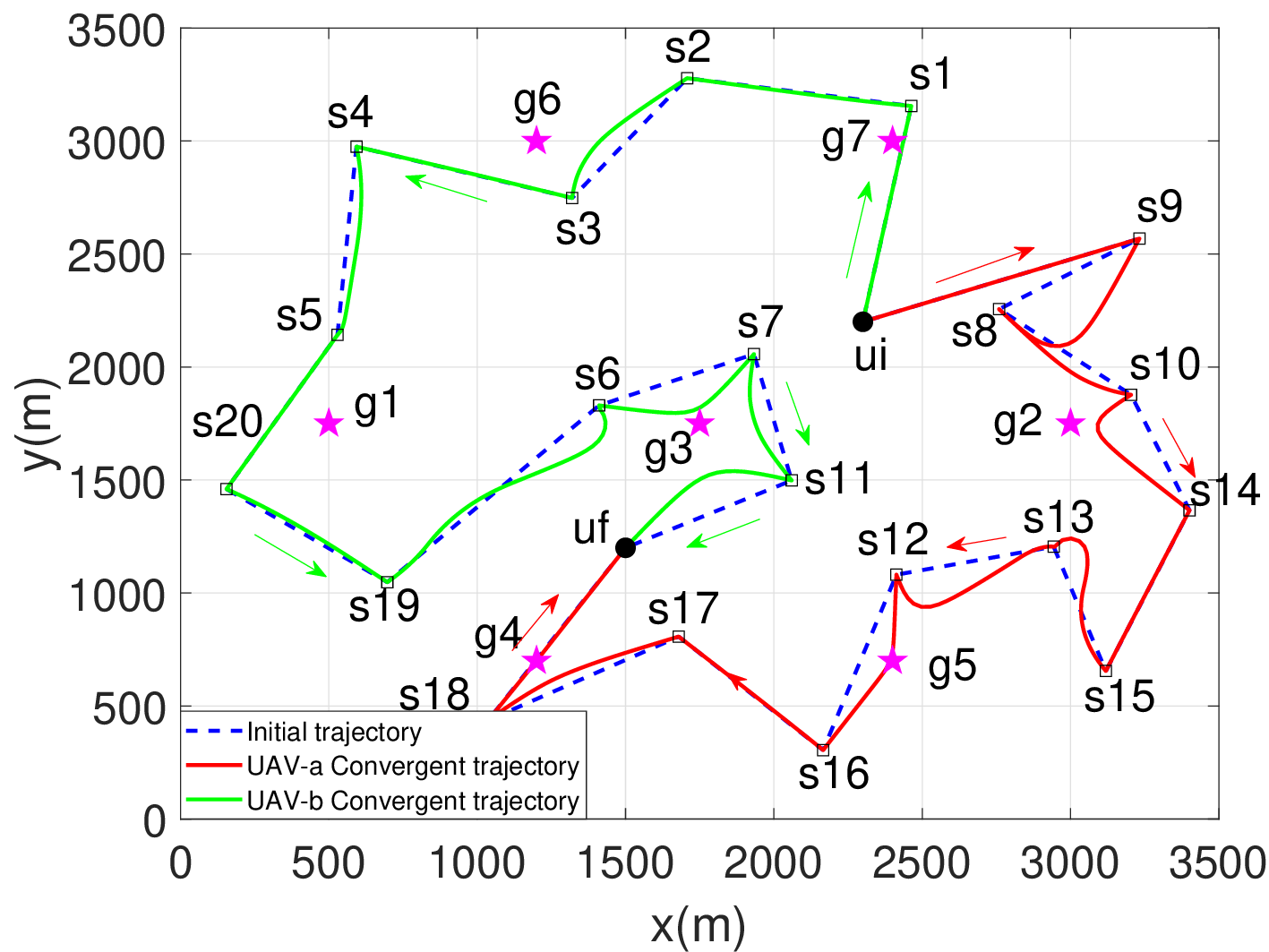}%
  \label{fig:right}}

  \vspace{1em}
  \scriptsize{$Q_{s_k}$=[208, 250, 212, 211, 52, 432, 247, 249, 396, 352, 356, 452, 300, 97, 252, 150, 312, 496, 459, 156] Mbits.}

  \caption{Optimized route of cruise points with two UAVs.}

  \label{fig:7}
\end{figure*}

\begin{figure}[!t]
  \centering
  \captionsetup[subfloat]{font=scriptsize} 
  
  \subfloat[Region division]{\includegraphics[width=1.5in]{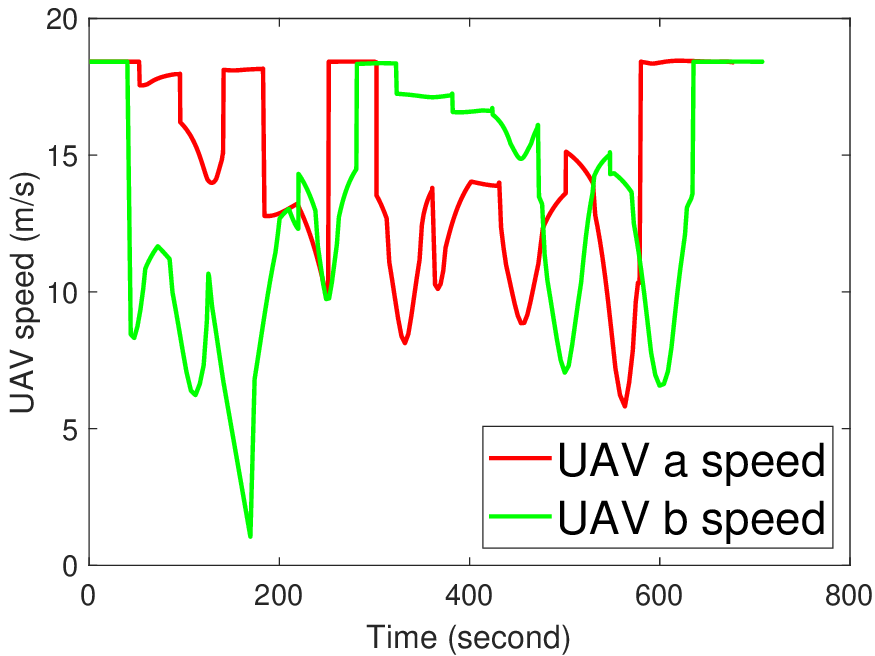}%
  \label{fig:6a}}
  \hfil
  \subfloat[Region division]{\includegraphics[width=1.5in]{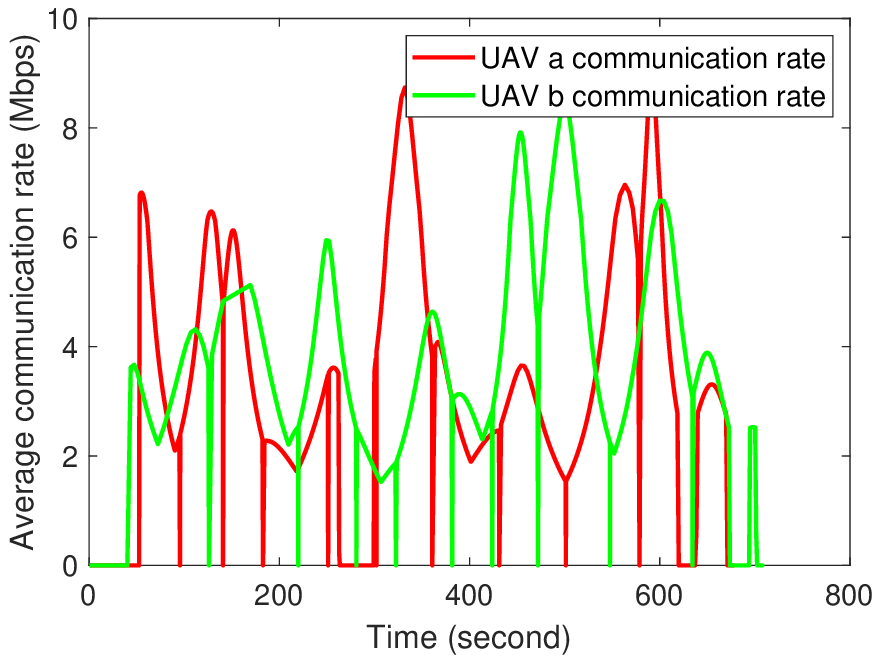}%
  \label{fig:6d}}
  \hfil
  
  \vspace{-1em}
  \captionsetup[subfloat]{justification=centering}
  \subfloat[Shortest distance criterion]{\includegraphics[width=1.5in]{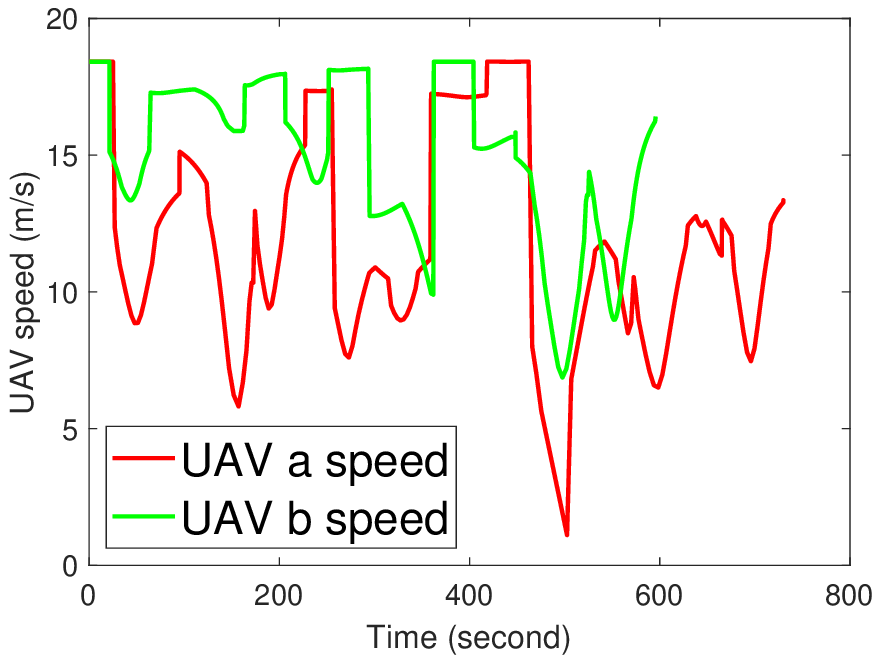}%
  \label{fig:6b}}
  \hfil
  \captionsetup[subfloat]{justification=centering}
  \subfloat[Shortest distance criterion]{\includegraphics[width=1.5in]{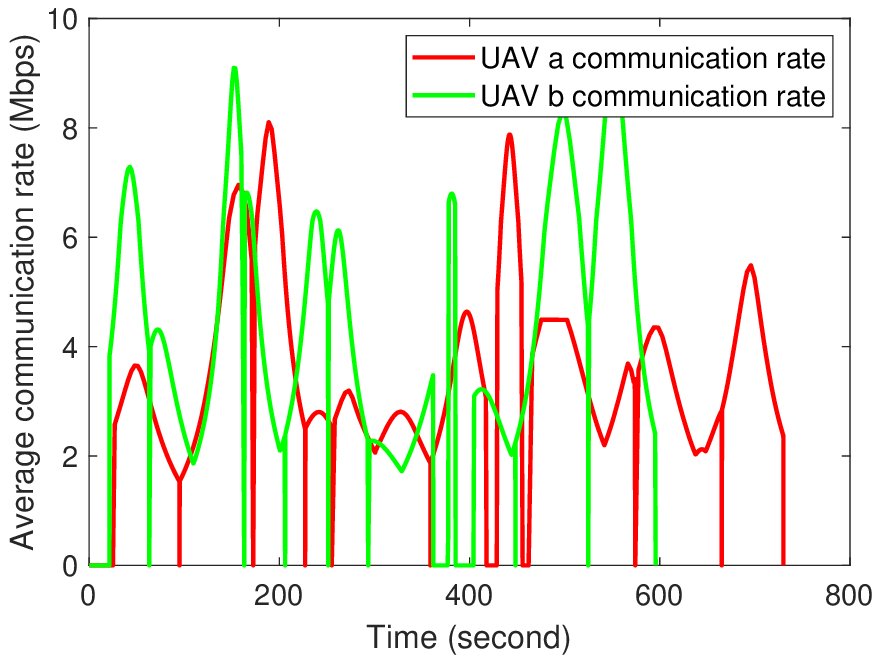}%
  \label{fig:6e}}
  \hfil

  \vspace{-1em}
  \subfloat[EBTAS]{\includegraphics[width=1.5in]{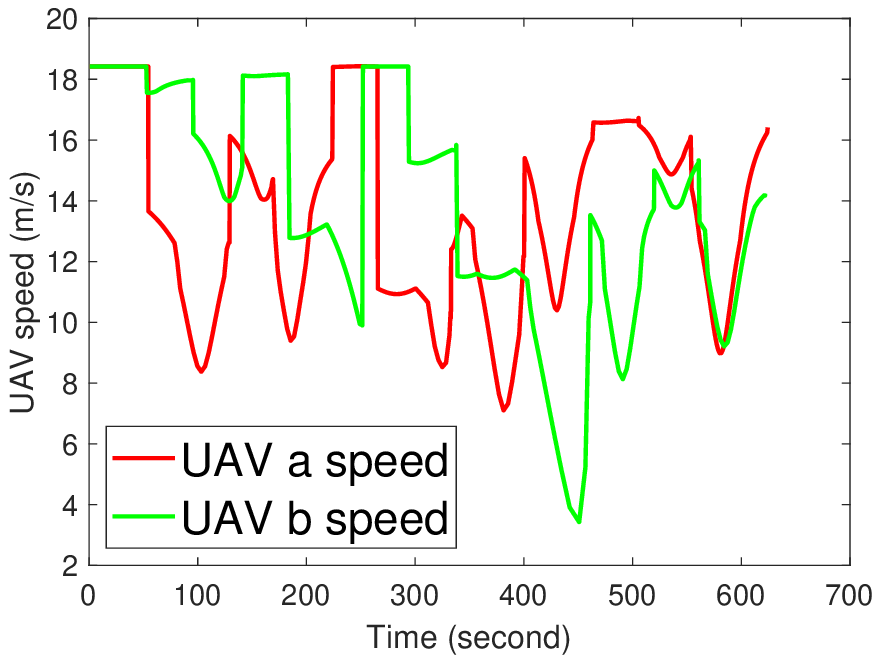}
  \label{fig:6c}}
  \hfil
  \subfloat[EBTAS]{\includegraphics[width=1.5in]{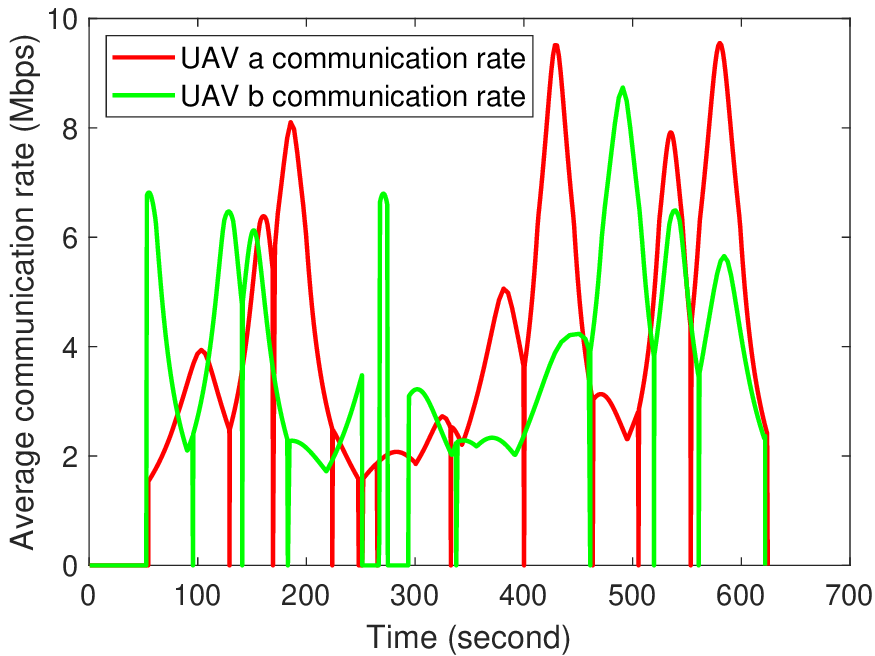}
  \label{fig:6f}}

  \tiny{$Q_{s_k}$=[208, 250, 212, 211, 52, 432, 247, 249, 396, 352, 356, 452, 300, 97, 252, 150, 312, 496, 459, 156] Mbits.}
  
  \captionsetup{justification=centering} 
  \caption{UAVs' speed and communication rate over time.}
  \label{fig:9}
\end{figure}

\begin{figure}[!t]
  \centering
  \includegraphics[width=2.5in]{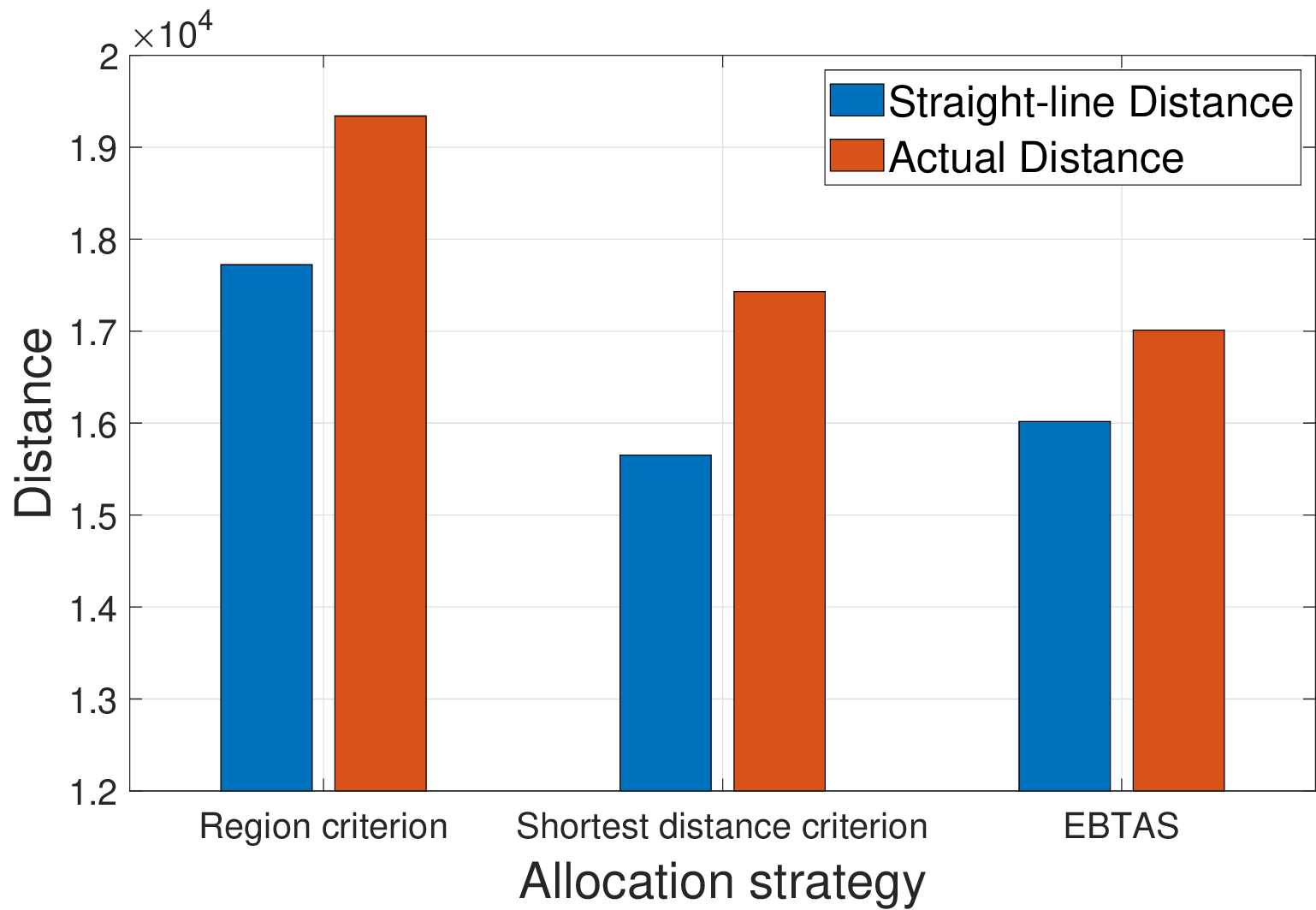}
  \caption{Comparison of UAV moving distance.}
  \label{fig:comparison}
  \end{figure}

Fig. \ref{fig:7} shows the optimized trajectory of the UAVs in the scenario where each cruise point has different communication offloading requirements. It can be seen from the figure that in the road sections with large computation, the UAV will not fly along the predetermined straight path, but will approach the direction of the base station, so as to obtain a large communication gain to offload the data to the base station for calculation. However, in this process, the UAV will generate additional flight energy consumption and communication energy consumption.

As shown in Fig. \ref{fig:7}(a), in the strategy based on geographic region division, although some cruise points have a large amount of communication data, the total moving distance of the UAV is long, and there is more time to process and offload data. Therefore, the UAV does not need to move too much additional distance to complete the calculation task, and more road sections fly straight. In Fig. \ref{fig:7}(b), in the scenario of high offload volume of cruise points, the UAV task allocation method based on the shortest moving path of the UAV has the shortest total distance to traverse the cruise point in a straight line. However, in the road segments with low communication rates or large offload volume, the UAV needs to move more extra distance to get closer to GBSs, and more flight and calculation energy consumption will be consumed in this process, such as $s_{12} - s_{11} $, $s_{11} - s_{17}$. As shown in Fig. 8(c), the task allocation algorithm based on the EBTAS shows advantages in this process, such as $s_{8}-s_{10}$ and $s_{11}-s_F$ segments. Although the distance between these segments is short and the offload volume is large, they are close to GBS $g_3$ and have strong signal gain, so they only need to move a small distance on the basis of a straight line to complete the calculation task. Thus, it is beneficial to reduce energy consumption.

\begin{table}[htbp]
  
    \caption{Performance comparison with two UAVs}
    \begin{center}
    \renewcommand{\arraystretch}{1.3} 
    \setlength{\tabcolsep}{3pt} 
    \tiny 
    \newcolumntype{C}[1]{>{\centering\arraybackslash}p{#1}} 
    \begin{tabular}{|C{2.4cm}|C{0.8cm}|C{0.8cm}|C{0.8cm}|C{0.8cm}|C{0.8cm}|C{0.8cm}|}
    \hline
    \multirow{2}{*}{Performance} & \multicolumn{2}{c|}{Region criterion} & \multicolumn{2}{c|}{Shortest distance criterion} & \multicolumn{2}{c|}{EBTAS}\\
    \cline{2-7}
     & UAV-a & UAV-b & UAV-a & UAV-b & UAV-a & UAV-b \\
    \hline
    Task completion time($s$) & 678.3 & 708.8 & 729.9 & 595.7 & 624.4 & 621.9 \\
    \cline{1-7}
    Energy consumption($J$) & 97134.5 & 100563.1 & 99439.1 & 84851.1 & 85641.5 & 87238.7 \\
    \hline
    Average time($s$) & \multicolumn{2}{c|}{693.6} & \multicolumn{2}{c|}{662.8} & \multicolumn{2}{c|}{623.2} \\
    \hline
    Total energy($J$) & \multicolumn{2}{c|}{197697.6} & \multicolumn{2}{c|}{184290.2} & \multicolumn{2}{c|}{172880.2} \\
    \hline
    \end{tabular}
    \label{tab:1}
    \end{center}
  \end{table}

\begin{figure*}[t]
  \centering
  \captionsetup[subfloat]{font=scriptsize} 
  \subfloat[Region criterion]{\includegraphics[width=2.3in]{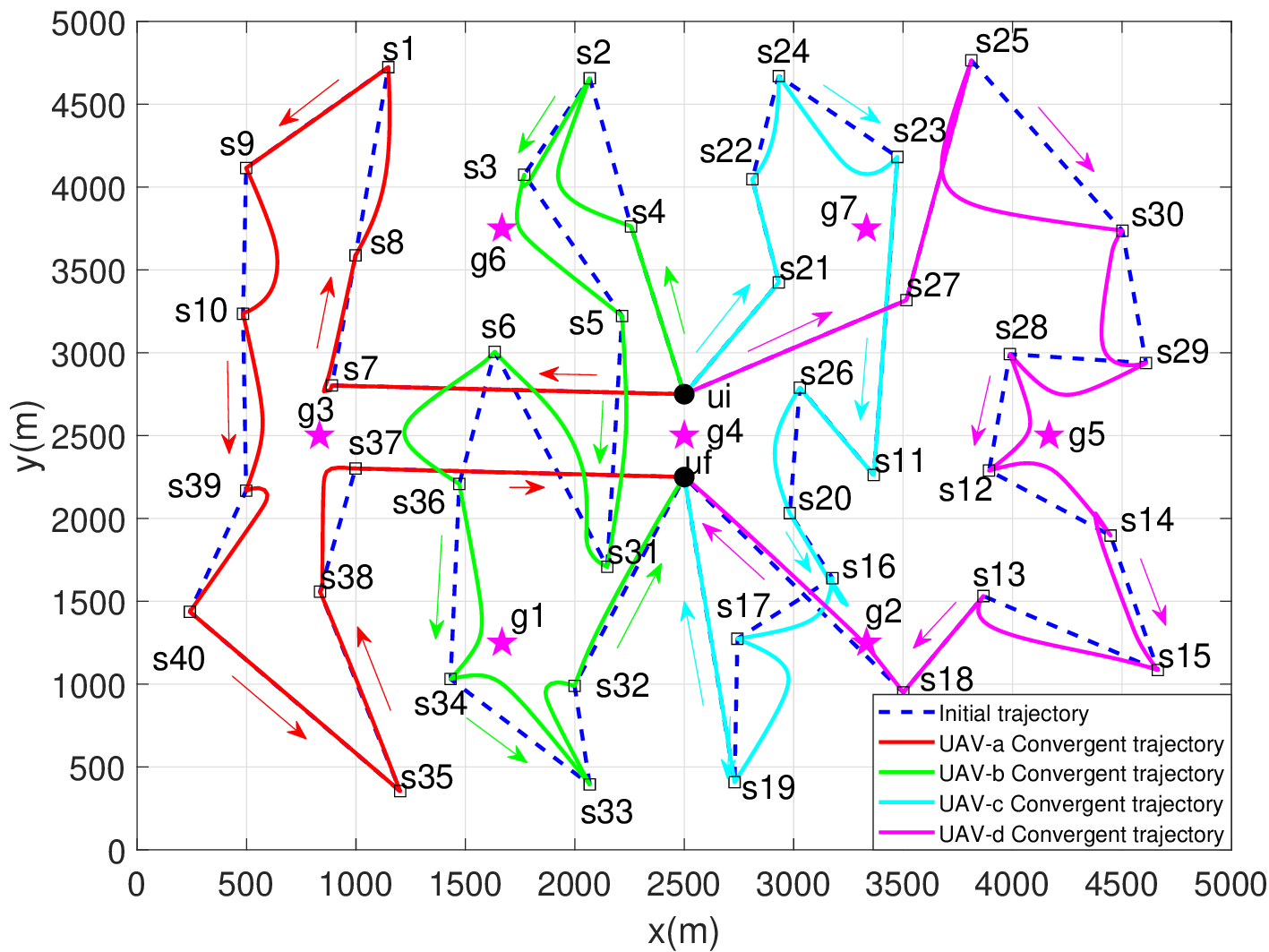}%
  \label{fig:left}}
  \hfil  
  \subfloat[Shortest distance criterion]{\includegraphics[width=2.3in]{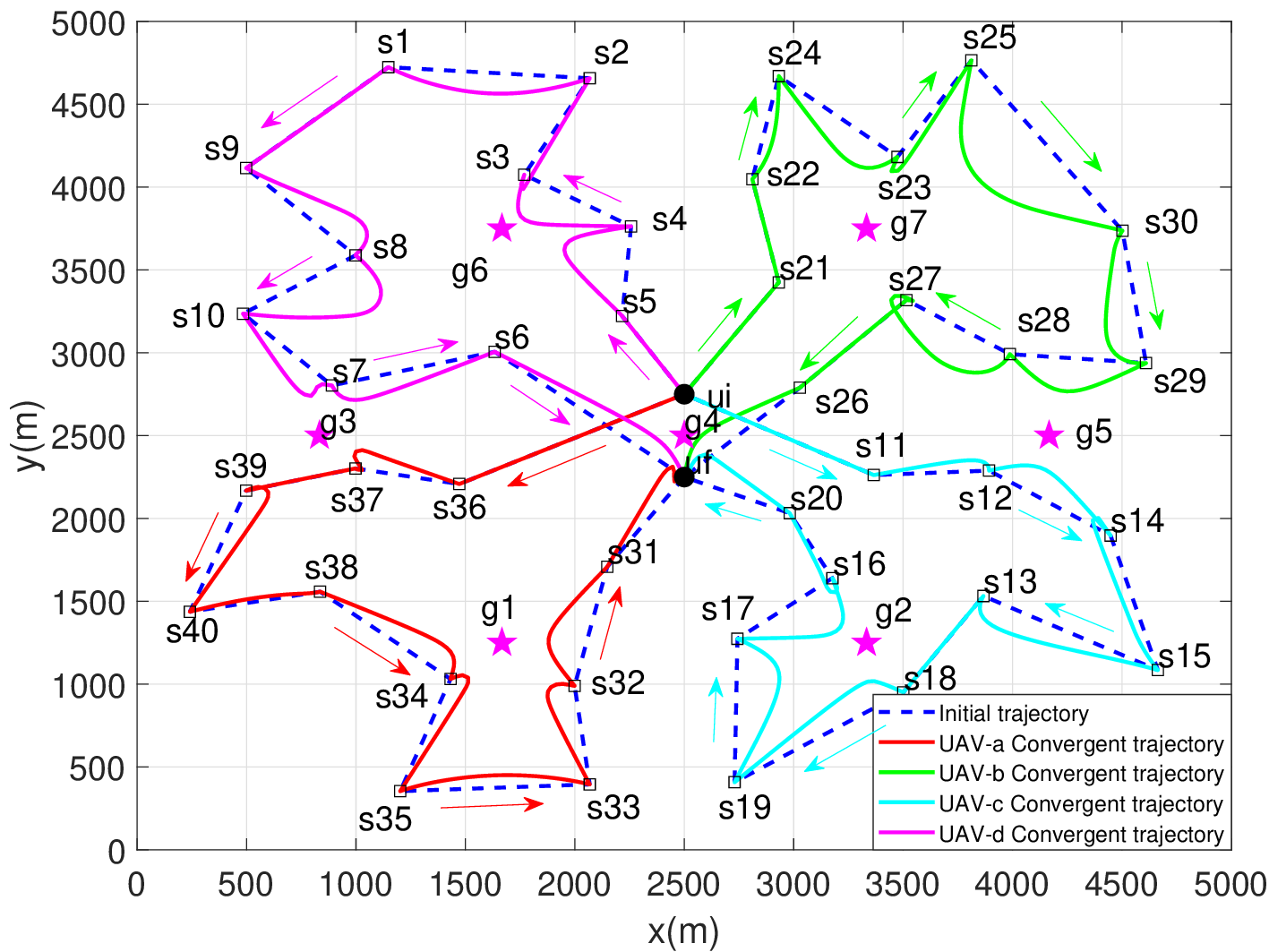}%
  \label{fig:center}}
  \hfil
  \subfloat[EBTAS]{\includegraphics[width=2.3in]{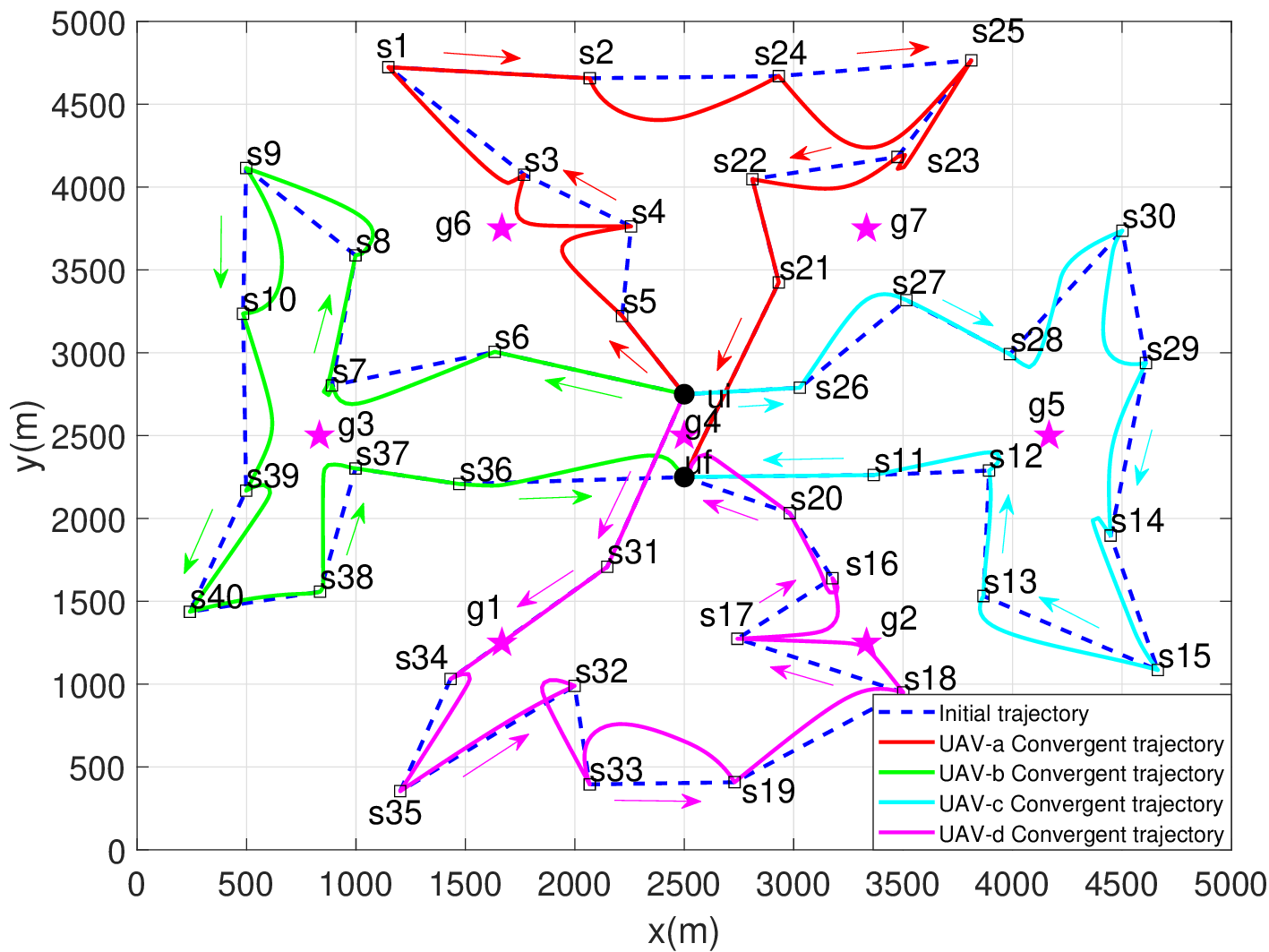}%
  \label{fig:right}}
  
  \tiny{$Q_{s_k}$ = [100, 212, 496, 459, 356, 352, 311, 212, 200, 500, 97, 347, 156, 452, 200, 232, 347, 552, 249, 396, 70, 123, 311, 359, 456, 252, 111, 412, 300, 228, 497, 307, 256, 352, 200, 432, 89, 372, 249,  96] Mbits.}
  \caption{Optimized route of cruise points with four UAVs.}

  \label{fig:10}
\end{figure*}

\begin{figure*}[t]
  \centering
  \captionsetup[subfloat]{font=scriptsize} 
  \subfloat[Region criterion]{\includegraphics[width=2.3in]{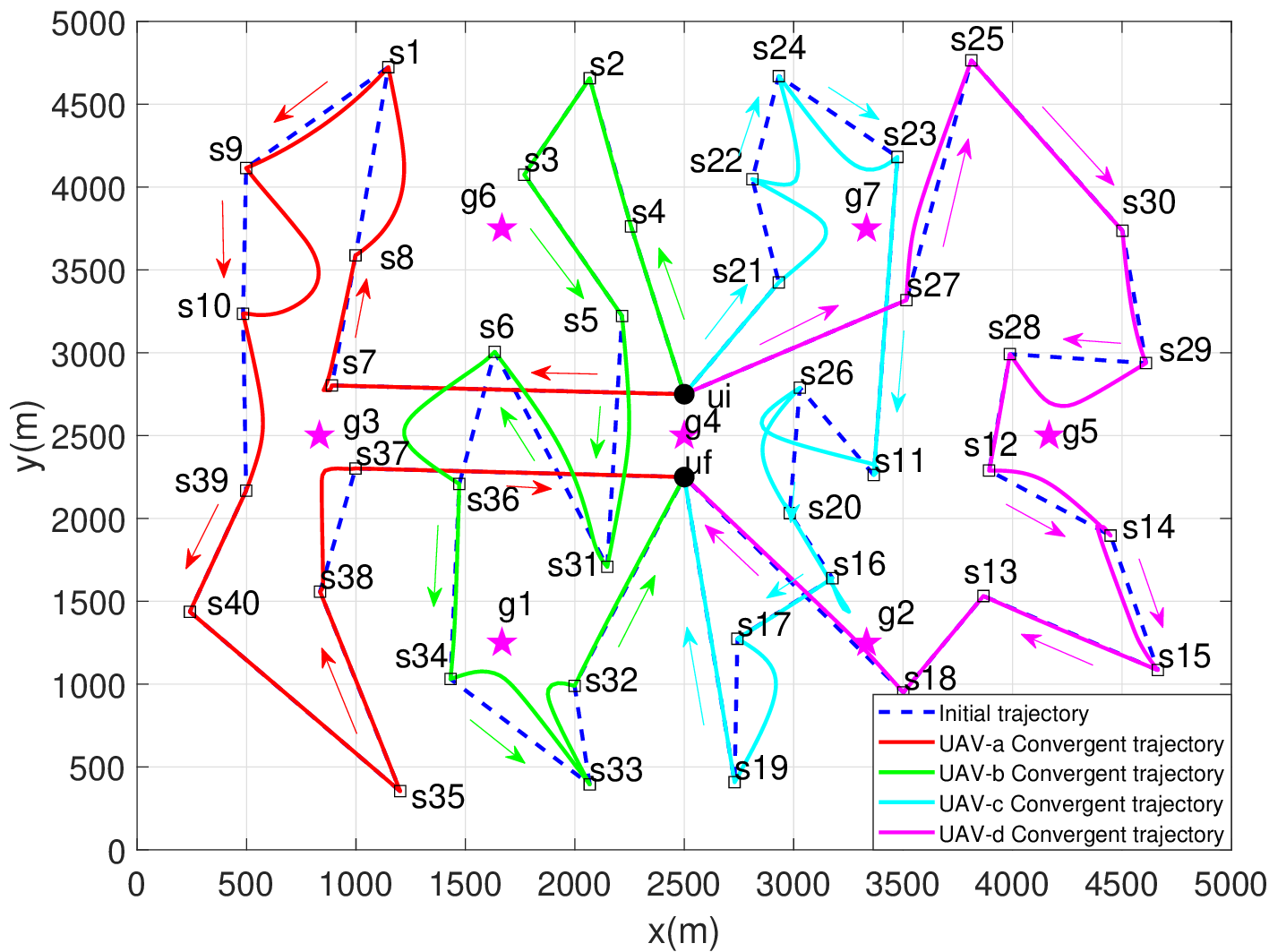}%
  \label{fig:left}}
  \hfil
  \subfloat[Shortest distance criterion]{\includegraphics[width=2.3in]{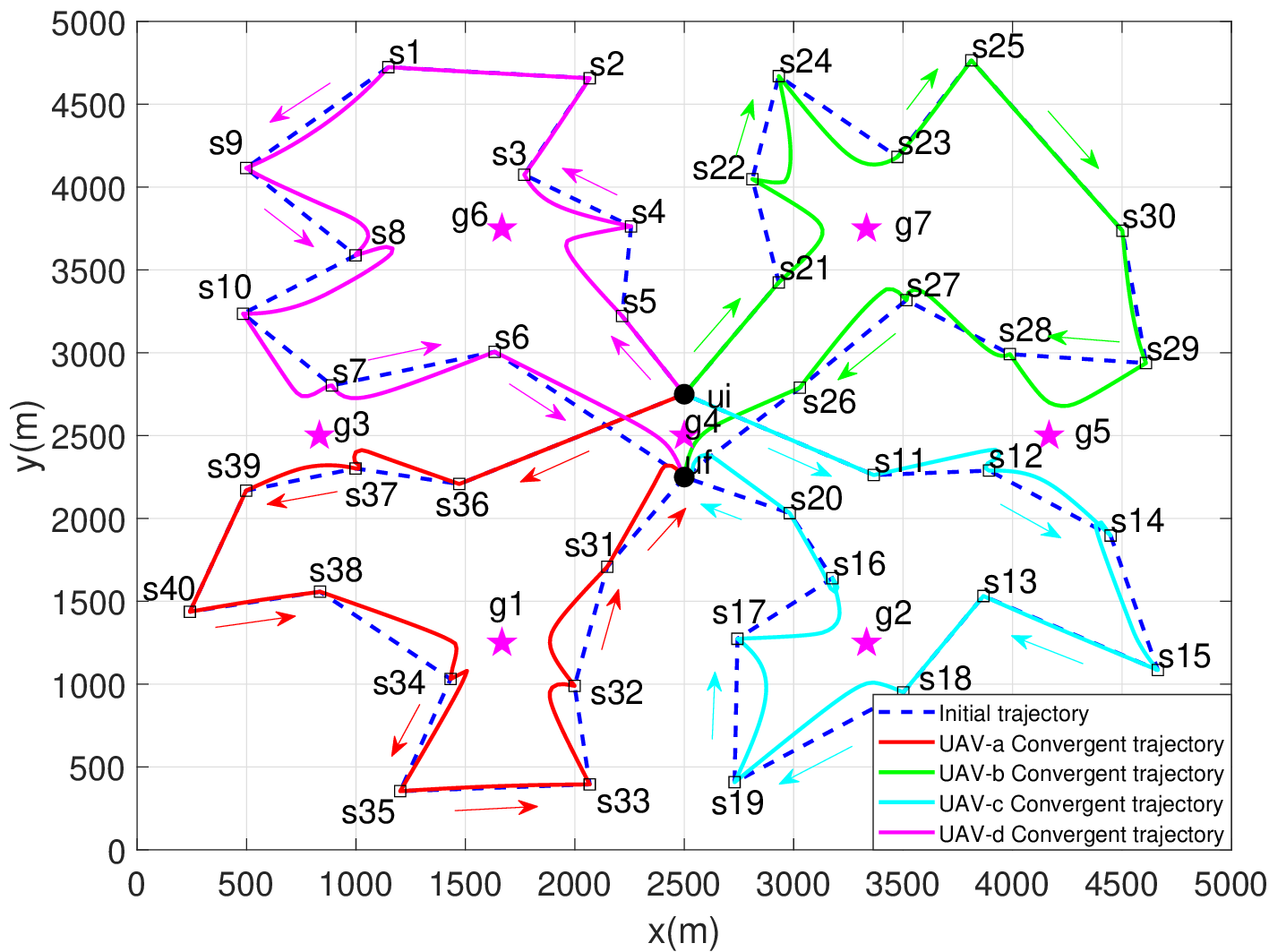}%
  \label{fig:center}}
  \hfil
  \subfloat[EBTAS]{\includegraphics[width=2.3in]{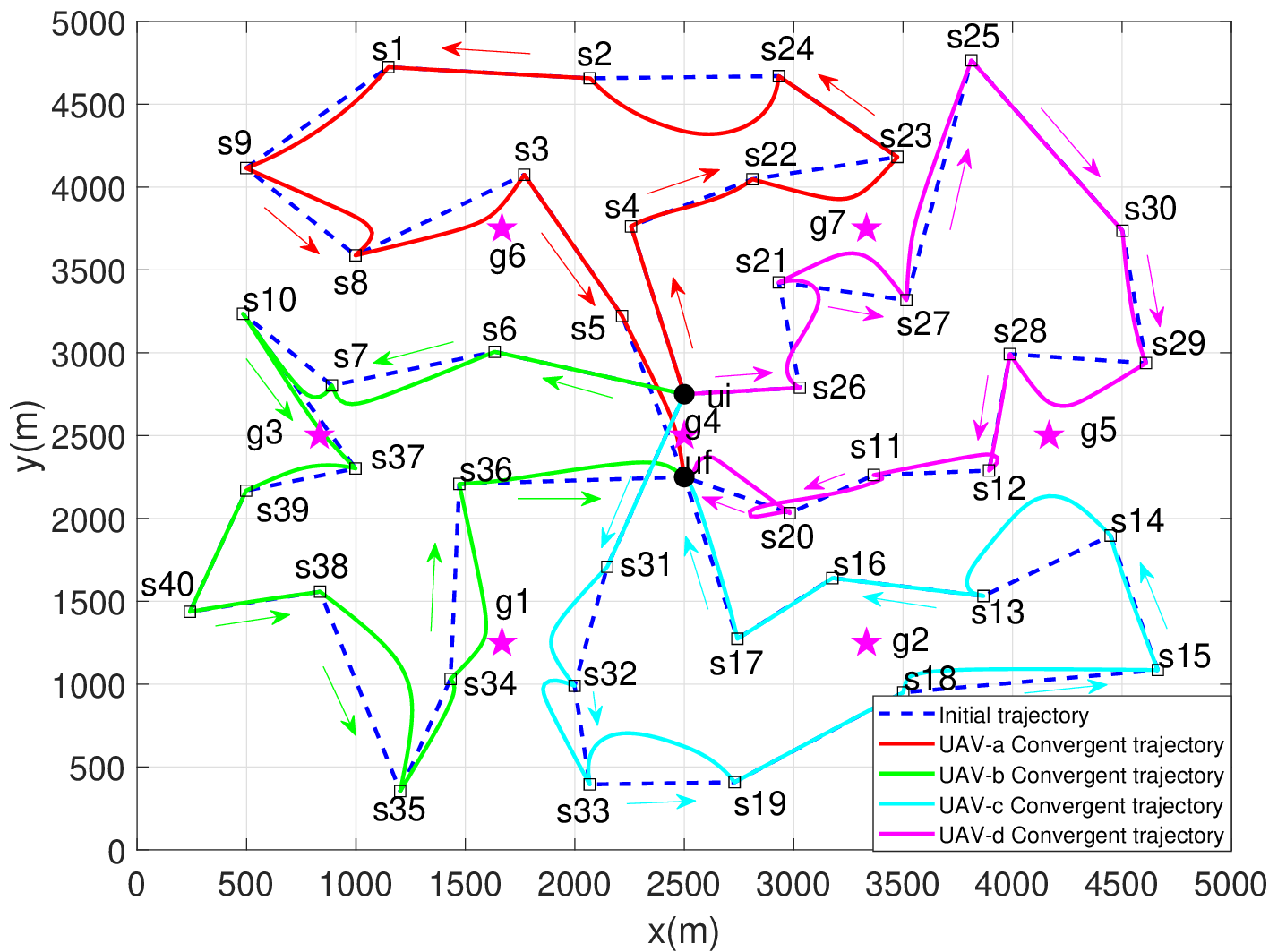}%
  \label{fig:right}}

  \tiny{$Q_{s_k}$ = [167, 109, 124, 140, 234, 154, 139, 351, 309, 119, 468, 377, 278, 409, 115, 135, 439, 192, 353, 474, 394, 287, 112, 252, 414, 290, 374, 206, 423, 318, 363, 380, 209, 489, 171,  85, 223, 447,  76,  81] Mbits.}

  \caption{Optimized route of cruise points with four UAVs.}

  \label{fig:11}
\end{figure*}
Combining the trajectory diagram and comparing Fig. \ref{fig:9}(a)-Fig. \ref{fig:9}(f), it can be seen that when the UAV moves between every two cruise points, the speed curve first decreases and then increases, while the communication rate curve first increases and then decreases. This indicates that the UAV will first approach the direction of GBS with a large speed to obtain a large communication rate in a computationally intensive scenario, then reduce its flight speed near GBS for data transmission, and finally accelerate to fly to the next cruise point.

\begin{table*}[htbp]
  \caption{Performance comparison with four UAVs}
  \begin{center}
  \renewcommand{\arraystretch}{1.3} 
  \setlength{\tabcolsep}{3pt} 
  \footnotesize 
  \newcolumntype{C}[1]{>{\centering\arraybackslash}p{#1}} 
  \begin{tabular}{|C{2.8cm}|C{1cm}|C{1cm}|C{1cm}|C{1cm}|C{1cm}|C{1cm}|C{1cm}|C{1cm}|C{1cm}|C{1cm}|C{1cm}|C{1cm}|}
  \hline
  \multirow{2}{*}{Performance} & \multicolumn{4}{c|}{Region criterion} & \multicolumn{4}{c|}{Shortest distance criterion} & \multicolumn{4}{c|}{EBTAS}\\
  \cline{2-13} 
   & UAV-a & UAV-b & UAV-c & UAV-d & UAV-a & UAV-b & UAV-c & UAV-d & UAV-a & UAV-b & UAV-c & UAV-d \\
  \hline
  Task completion time($s$) & 991 & 837.9 & 889.8 & 786.3 & 790.6 & 816.4 & 809.2 & 870.8 & 789.8 & 737.1 & 800.2 & 766.8 \\
  \cline{1-13}
  Energy consumption($J$) & 140438 & 118521.5 & 125175.2 & 116515.3 & 110513.4 & 113701.8 & 113918.1 & 120501.4 & 109038.4 & 101479.6 & 110688.7 & 105876.3 \\
  \hline
  Average time($s$) & \multicolumn{4}{c|}{876.2} & \multicolumn{4}{c|}{821.8} & \multicolumn{4}{c|}{773.5} \\
  \hline
  Total energy($J$) & \multicolumn{4}{c|}{500650} & \multicolumn{4}{c|}{458634.7} & \multicolumn{4}{c|}{427083} \\
  \hline
  \end{tabular}
  \label{tab:2}
  \end{center}
\end{table*}

Fig. \ref{fig:comparison} compares the straight-line distances and actual distances traveled by UAVs under three different allocation strategies. It is observed that the allocation strategy based on the shortest path results in the minimum straight-line travel distance for the UAVs. However, since UAVs need to complete data offloading tasks, they do not move in a straight line most of the time. In contrast, the EBTAS allocation strategy tends to select segments with better communication rates for task execution. As a result, the UAV does not need to travel much additional distance to complete data transmission tasks, resulting in a shorter actual travel distance.

Table \ref{tab:1} presents the energy and time consumption metrics for each UAV under three partitioning methods. It is evident from the table that the proposed balanced task allocation strategy outperforms the traditional partitioning in terms of task completion time and energy consumption. Notably, under the EBTAS strategy, the task completion times and energy consumption of the two UAVs also exhibited smaller variances, suggesting an efficient utilization of the wireless resources in the patrol area for data transmission, while ensuring a balanced task allocation considering the load of each UAV.

In consideration of more intricate scenarios involving four UAVs and forty cruise points, Fig. \ref{fig:10} illustrates the optimized patrol trajectories of UAVs under the communication offload volume conditions depicted in the figure. As shown in Fig.\ref{fig:10}(a), UAVs are allocated tasks based on geographic regions, with the patrol area evenly divided into four sections assigned to each UAV. Fig.\ref{fig:10}(b) depicts task allocation based on the shortest straight-line movement distance for UAVs, while Fig.\ref{fig:10}(c) is based on the EBTAS method for UAV task assignment. In the same patrol scenario, as the communication offloading demand of the cruise points varies, the optimized paths of UAVs are illustrated in Fig. \ref{fig:11}. A comparison with Fig. \ref{fig:10} reveals that task allocation strategies based on geographic regions or UAVs' shortest straight-line movement paths do not make corresponding adjustments, whereas EBTAS dynamically allocates tasks based on the actual communication offload volume, ensuring efficient and balanced completion of patrol tasks by each UAV. Table \ref{tab:2} presents a performance comparison of the three task allocation strategies corresponding to the patrol scenarios in Fig. \ref{fig:11}, demonstrating that the EBTAS allocation strategy outperforms traditional strategies in terms of average task completion time and energy consumption. The numerical results demonstrate that the proposed algorithm possesses high adaptability and practicality in patrol scenarios of varying scales.

\section{Conclusion and Future Work}\label{SCM}
In this paper, we propose an energy-efficient and balanced task assignment strategy for the cellular-connected multi-UAV patrol scenario, which allocates patrol tasks to UAVs based on factors such as geographical location, offloading amount, and communication intensity. Numerical results show that, compared to traditional task assignment algorithms, the proposed method offers advantages in terms of task completion time and energy consumption. It can flexibly adapt to patrol areas of different sizes and varying task offloading demands, demonstrating significant practical application potential. In future work, with the rapid advancement of deep learning and AI methods, further research could explore integrating radio map platforms with these techniques to enhance performance for the given scenario. This approach may lead to more adaptive and intelligent task assignment solutions.

\bibliographystyle{IEEEtran}
\bibliography{IEEEabrv,reference}

\newpage

\section{Biography Section}

\vspace{-3.5em}

\begin{IEEEbiography}[{\includegraphics[width=1in,height=1in,clip,keepaspectratio]{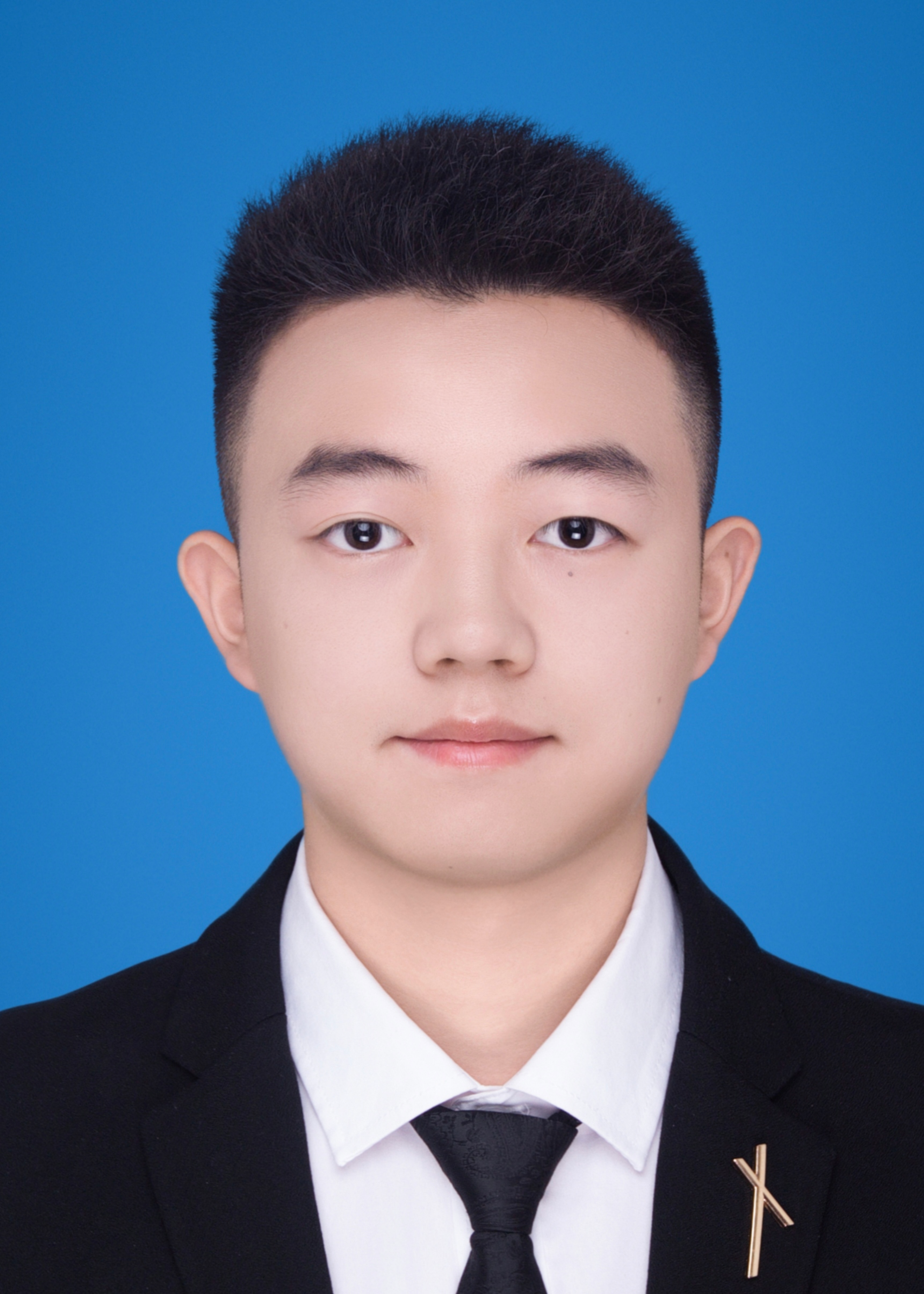}}]{Kuan Jia}
is currently pursuing a Bachelor's degree in the School of Information Engineering, Nanchang University. His research interests include MEC technology, UAV communication and wireless resource management.
\end{IEEEbiography}\vspace{-4em}
\begin{IEEEbiography}[{\includegraphics[width=1in,height=1in,clip,keepaspectratio]{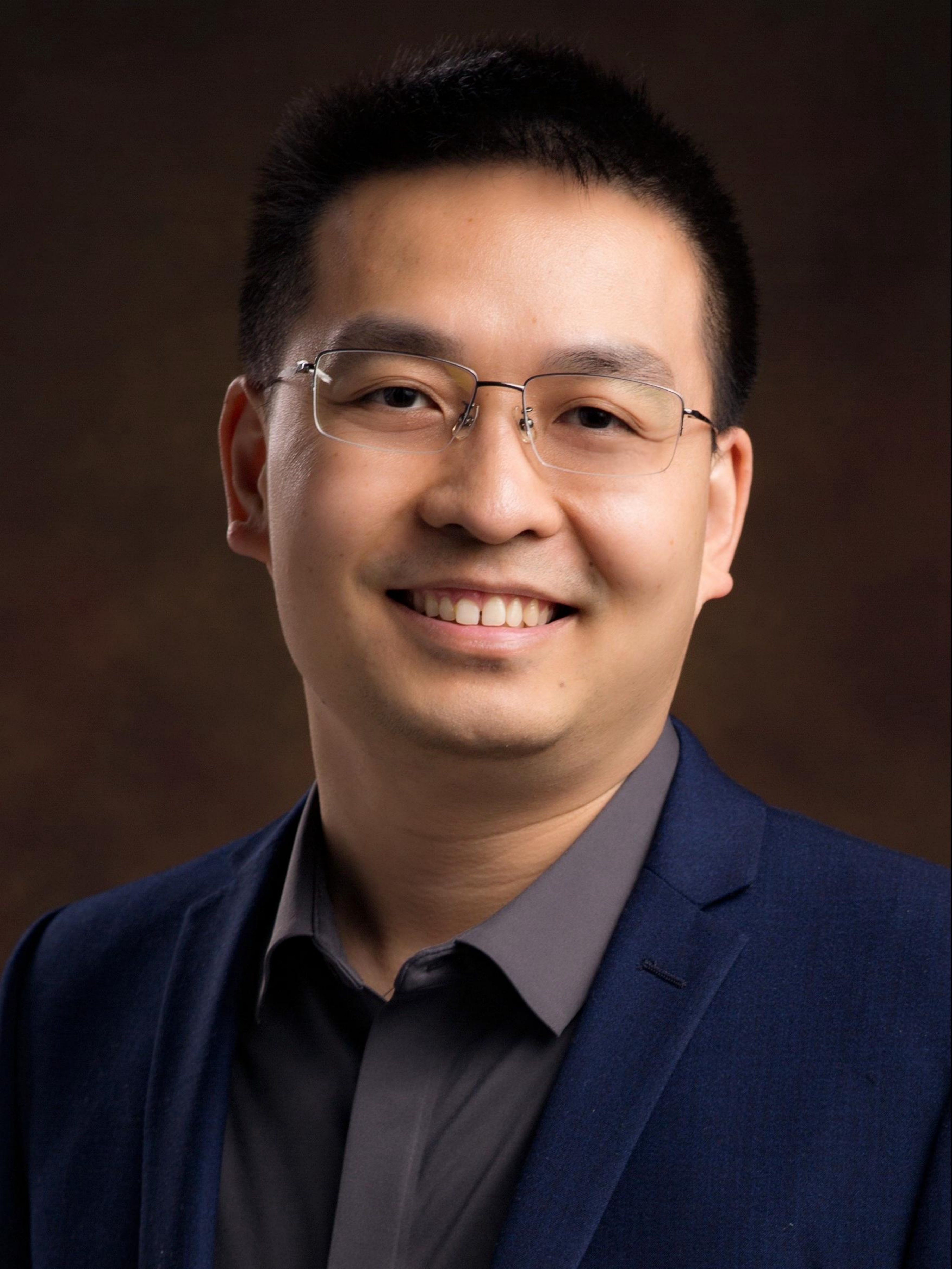}}]{Dingcheng Yang}
received the B.S. degree in electronic engineering and the Ph.D. degree in space
physics from Wuhan University, Wuhan, China, in 2006 and 2012, respectively. He is currently a Professor with the Information Engineering School, Nanchang University, Nanchang, China. He had published more than 50 papers including journal papers on IEEE TRANSACTIONS
on VEHICULAR TECHNOLOGY, etc. and conference papers such as IEEE GLOBECOM. His research interests include cooperation communications, IoT/cyberphysical systems, UAV communications, and wireless resource management.  
\end{IEEEbiography}\vspace{-3em}
\begin{IEEEbiography}[{\includegraphics[width=1in,height=1in,clip,keepaspectratio]{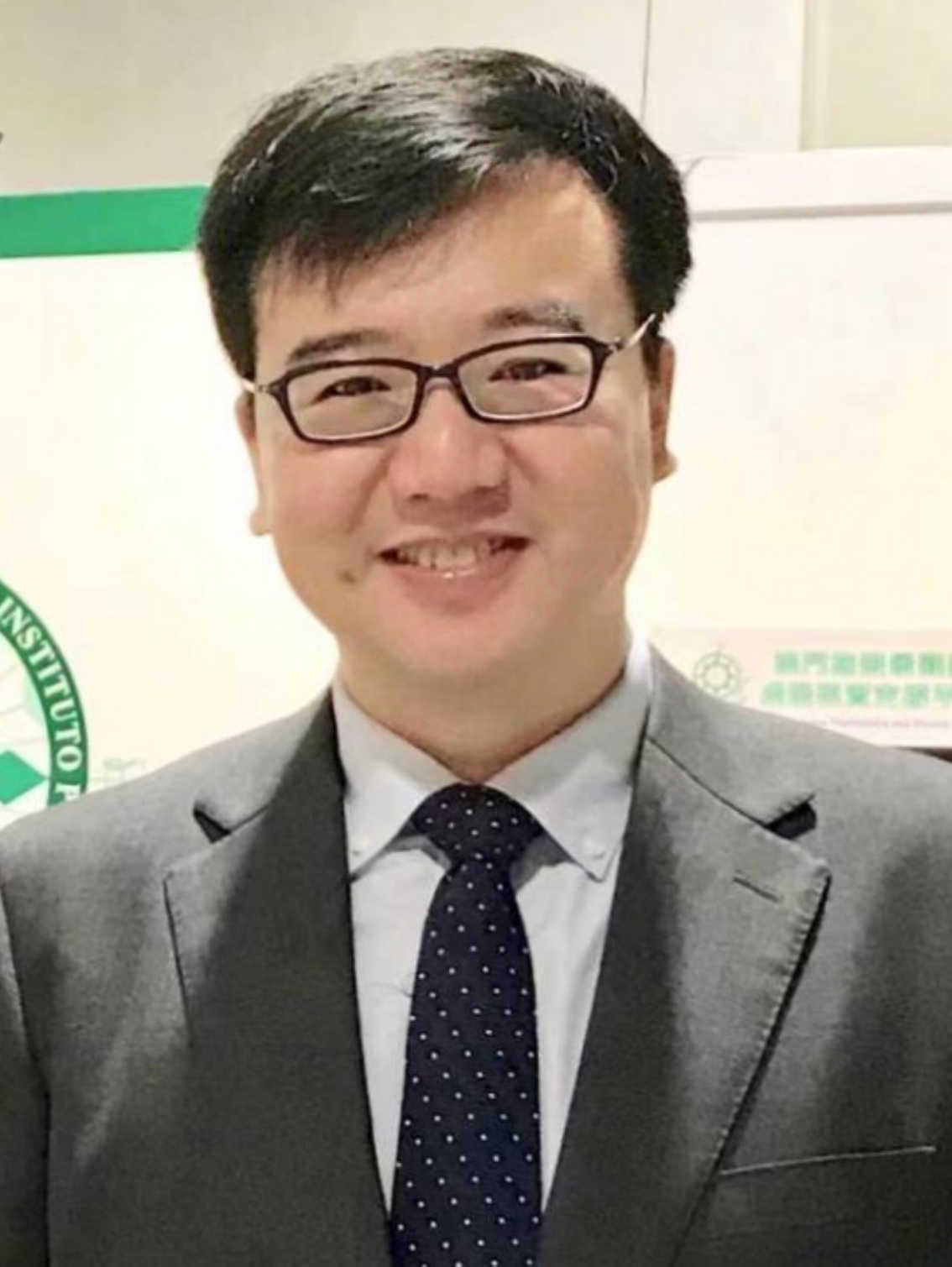}}]{Yapeng Wang}
  received the B.Sc. in North China Electric Power University, China in 1998, M.Sc. and Ph.D. degrees in Queen Mary University of London, UK in 2002 and 2007. He joined the Faculty of Applied Sciences, Macao Polytechnic University in 2021 as an associate professor. His current research interests include wireless communications, automatic speech recognition, machine learning.  
\end{IEEEbiography}\vspace{-4em}
\begin{IEEEbiography}[{\includegraphics[width=1in,height=1in,clip,keepaspectratio]{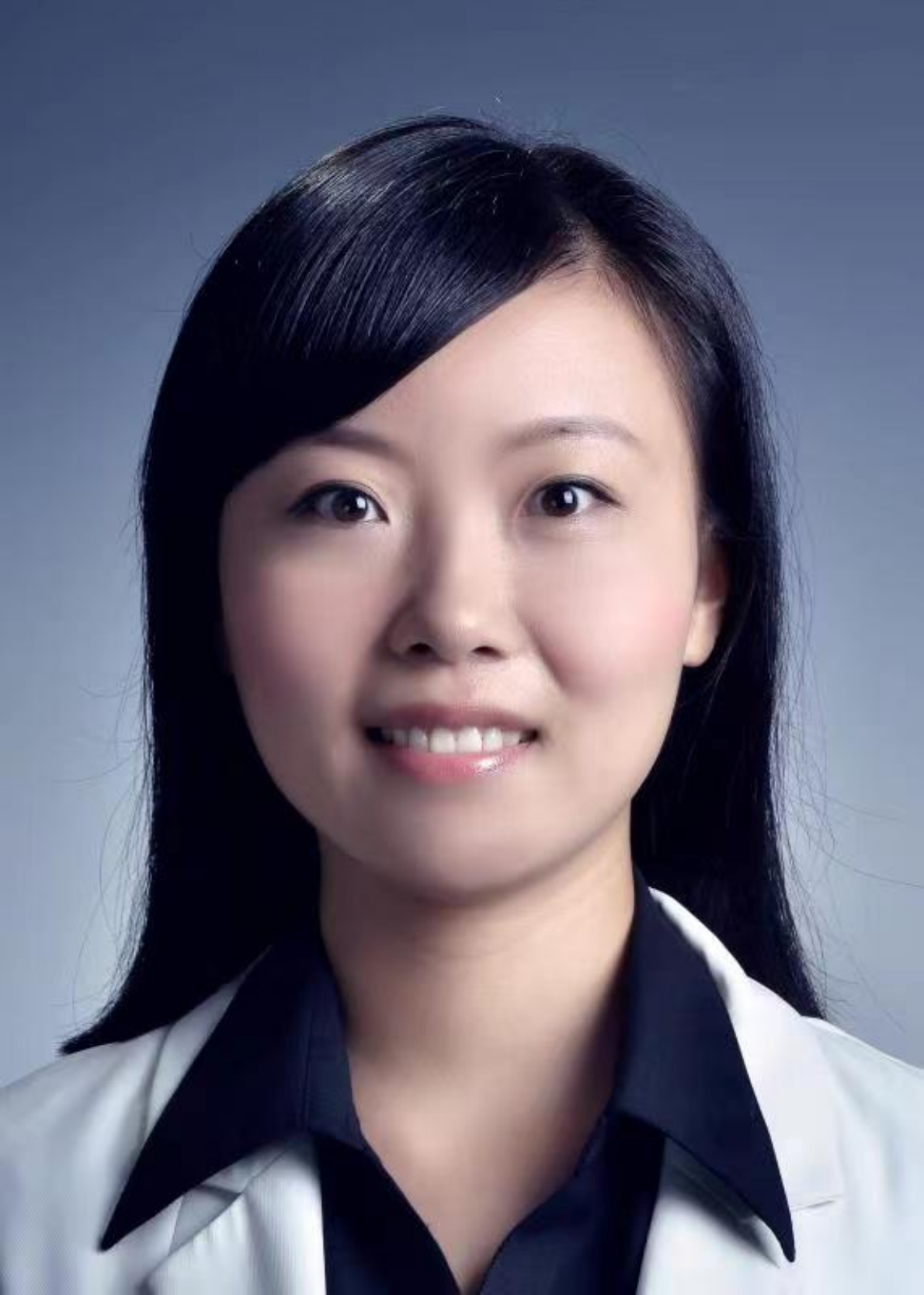}}]{Tianyun Shui}
  received the B.S.degree in communications engineering and the M.S.degree in electronic engineering from University of Electronic Science and Technology of China, Chengdu,China,in 2011 and 2014,respectively. Her research interests include big data and radio network communication.  
\end{IEEEbiography}\vspace{-4em}
\begin{IEEEbiography}[{\includegraphics[width=1in,height=1in,clip,keepaspectratio]{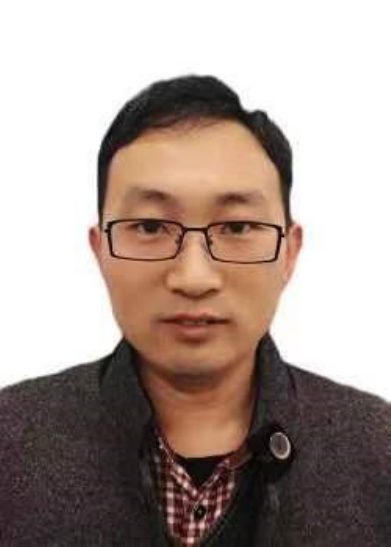}}]{Chenji Liu}
  received the M.S degree from college of information science and engineering, Zhejiang University,Hangzhou, China, in 2004,then worked as a Senior Software Systems Engineer at ZTE R\&D Center,Shanghai,China, for 10 years.Currently serves as a R\&D leader at Wireless Network Department of China Mobile Communications Group Jiangxi Co., Ltd  
\end{IEEEbiography}\vspace{-4em}

\vspace{11pt}

\vfill

\end{document}